\documentclass{statsoc}

\usepackage[final]{graphicx}
\usepackage{amsmath,amsfonts,amssymb}
\usepackage{verbatim}
\usepackage{color}
\usepackage{float}
\usepackage[draft=false,letterpaper,breaklinks,colorlinks,linktocpage,citecolor=blue,linkcolor=blue,urlcolor=blue]{hyperref}

\newcommand{\BF}[1]{\mbox{\boldmath$#1$}}

\newcommand{\DP}[2]{\ensuremath{\mathrm{DP}\!\left({#1},{#2}\right)}}

\newcommand{\MN}[4]{\mathcal{M}\mathcal{N}\left(#1;#2,#3,#4\right)}

\newcommand{\symsubsup}[3]{{#1}_{#2}^{(#3)}}
\newcommand{\symsubsupB}[3]{\mathbf{#1}_{#2}^{(#3)}}
\newcommand{\symsubsupT}[3]{\mathbf{#1}_{#2}^{(#3)^T}}

\newcommand{\newfeatures}[1]{n_{#1}}
\newcommand{\Kplus}{\ensuremath{K_{\!+}}}
\newcommand{\fset}[1]{\ensuremath{\boldsymbol{f}_{\!#1}}}

\newcommand{\figdir}{figs}

\floatstyle{boxed}
\newfloat{algorithm}{tbh}{loa}
\floatname{algorithm}{Algorithm}
\newcommand{\algtop}{
  \vspace*{-6pt}
  \setlength{\belowdisplayskip}{2pt plus 2pt}
  \setlength{\abovedisplayskip}{2pt plus 2pt}
  \setlength{\itemsep}{4pt}
}
\newcommand{\algend}{
  \vspace*{-2pt}
}

\title{Joint Modeling of Multiple Related Time Series via the Beta Process} 
\author{Emily Fox} \address{University of Pennsylvania, Philadelphia, PA USA.} \email{ebfox@wharton.upenn.edu} 
\author{Erik Sudderth} \address{Brown University, Providence, RI USA.} \email{sudderth@cs.brown.edu} 
\author{Michael I. Jordan} \address{University of California, Berkeley, CA USA.} \email{jordan@stat.berkeley.edu} 
\author[Fox et. al.]{Alan Willsky} \address{Massachusetts Institute of Technology, Cambridge, MA USA.} \email{willsy@mit.edu}

\begin{document}

\begin{abstract}
	We propose a Bayesian nonparametric approach to the problem of jointly modeling multiple related time series.  Our approach is based on the discovery of a set of latent, shared dynamical behaviors.  Using a beta process prior, the size of the set and the sharing pattern are both inferred from data.  We develop efficient Markov chain Monte Carlo methods based on the Indian buffet process representation of the predictive distribution of the beta process, without relying on a truncated model.  In particular, our approach uses the sum-product algorithm to efficiently compute Metropolis-Hastings acceptance probabilities, and explores new dynamical behaviors via birth and death proposals. We examine the benefits of our proposed feature-based model on several synthetic datasets, and also demonstrate promising results on unsupervised segmentation of visual motion capture data. 
\end{abstract}


\section{Introduction}
Classical time series analysis has generally focused on a single (potentially multivariate) time series from which inferences are to be made.  For example, one might monitor the daily returns of a particular stock index and wish to infer the changing regimes of volatility.  However, in a growing number of fields, interest is in making inferences based on a \emph{collection} of related time series.  One might monitor multiple financial indices, or collect EEG data from a given patient at multiple non-contiguous epochs.  We focus on time series with dynamics that are too complex to be described using standard linear dynamical models (e.g., autoregressive processes), but that exhibit switches among a set of \emph{behaviors} that describe locally coherent and simple dynamic modes that persist over a segment of time.  For example, stock returns might be modeled via switches between regimes of volatility or an EEG recording between spiking patterns dependent on seizure type.  In such cases, one would like to discover and model the dynamical behaviors which are shared among several related time series.  In essence, we would like to capture a combinatorial form of shrinkage involving subsets of behaviors from an overall library of behaviors.

As a specific motivating example that we consider later in this paper, consider a multivariate time series that arises when position and velocity sensors are placed on the limbs and joints of a person who is going through an exercise routine.  In the specific dataset that we analyze, the time series can be segmented into types of exercise (e.g., jumping jacks, touch-the-toes and twists).  The goal is to discover these exercise types (i.e., the ``behaviors'') and their occurrences in the data stream.  Moreover, the overall dataset consists of multiple time series obtained from multiple individuals, each of whom performs some subset of exercise types.  We would like to take advantage of the overlap between individuals, such that if a ``jumping-jack behavior'' is discovered in the time series for one individual then it can be used in modeling the data for other individuals.

A flexible yet simple method of describing single time series with such patterned behaviors is the class of \emph{Markov switching processes}.  These processes assume that the time series can be described via Markov transitions between a set of latent dynamic behaviors which are individually modeled via temporally independent or linear dynamical systems.  Examples include the hidden Markov model (HMM), switching vector autoregressive (VAR) process, and switching linear dynamical system (SLDS).  These models have proven useful in such diverse fields as speech recognition, econometrics, neuroscience, remote target tracking, and human motion capture.  In this paper, we focus our attention on the descriptive yet computationally tractable class of switching VAR processes.  In this case, the state, or \emph{dynamical mode}, of the underlying Markov process encodes the dynamic behavior exhibited at a given time step and each dynamic behavior is a VAR process.  That is, conditioned on the Markov-evolving state, the likelihood is simply a VAR model.

To discover the dynamic behaviors shared between multiple time series, we propose a feature-based model.  Globally, the collection of time series can be described by the shared \emph{library} of possible dynamic behaviors.  Individually, however, a given time series will only exhibit some subset of these behaviors.  That is, each time series has a \emph{vocabulary} of possible states.  The goal in relating the time series is to discover which behaviors are shared amongst the time series and which are unique.  Let us represent the vocabulary of time series $i$ by a \emph{feature vector} $\fset{i}$, with $f_{ik}=1$ indicating that time series $i$ has behavior $k$ in its vocabulary.  We seek a prior for these feature vectors.  We particularly aim to allow flexibility in the number of total and time-series-specific behaviors, and to encourage time series to share similar subsets of the large set of possible behaviors.  Our desiderata motivate a feature-based Bayesian nonparametric approach based on the \emph{beta process}~\citep{Hjort:90,Thibaux:07}.  Such an approach allows for \emph{infinitely} many potential dynamic behaviors, but encourages a sparse representation.

In our scenario, one can think of the beta process as defining a coin-flipping probability for each of an infinite set of possible dynamic behaviors.  Each time series' feature vector is modeled as the result of a Bernoulli process draw: the beta-process-determined coins are flipped for each dynamic behavior and the set of resulting heads indicate the set of selected features (implicitly defining an infinite-dimensional feature vector.)  The properties of the beta process induce sparsity in the feature space by encouraging sharing of features among the Bernoulli process observations.  Specifically, the total sum of coin weights is finite and only certain dynamic behaviors have large coin weights.  Thus, certain dynamic behaviors are more prevalent in the vocabularies of the time series, though the resulting vocabularies clearly need not be identical.  As shown by~\citet{Thibaux:07}, integrating over the latent beta process random measure (i.e., coin-flipping weights) induces a predictive distribution on features known as the \emph{Indian buffet process} (IBP)~\citep{GriffithsGhahramani:05}.  Computationally, this representation is key.  Given a sampled feature set, our model reduces to a collection of finite Bayesian VAR processes with partially shared parameters. 

Our presentation is organized as follows.  The beta process is reviewed in Section~\ref{background:BetaProcess}, following a brief overview of Markov switching processes.  In Section~\ref{sec:model}, we present our proposed beta-process-based model for jointly modeling multiple related Markov switching processes.  Efficient posterior computations based on a Markov chain Monte Carlo (MCMC) algorithm are developed in Section~\ref{sec:MCMC}.  The algorithm does not rely on model truncation; instead, we exploit the finite dynamical system induced by a fixed set of features to efficiently compute acceptance probabilities, and reversible jump birth and death proposals to explore new features.  The sampling of features relies on the IBP interpretation of the beta process---the connection between the beta process and the IBP is outlined in Section~\ref{sec:IBP}.  In Section~\ref{sec:related}, we describe related approaches.  Section~\ref{sec:synth} examines the benefits of our proposed feature-based model on several synthetic datasets.  Finally, in Section~\ref{sec:MoCap} we present promising results on the challenging task of unsupervised segmentation of data from the CMU motion capture database~\citep{CMUmocap}.
\section{Background}
\subsection{Markov Switching Processes}
\label{sec:MarkovSwitchingProcesses}
\subsubsection*{Hidden Markov Models}
The hidden Markov model, or \emph{HMM}, is a class of doubly
stochastic processes based on an underlying, discrete-valued state
sequence that is modeled as Markovian~\citep{Rabiner:89}. Conditioned
on this state sequence, the model assumes that the observations,
which may be discrete or continuous valued, are independent. Specifically,
let $z_t$ denote the state, or \emph{dynamical mode}, of the Markov chain at time~$t$
and let $\pi_j$ denote the state-specific \emph{transition distribution} for
mode $j$. Then, the Markovian structure on the mode sequence
dictates that
\begin{align}
z_t\mid z_{t-1} \sim \pi_{z_{t-1}}. \label{eqn:HMMmode}
\end{align}
Given the mode $z_t$, the observation $y_t$ is a conditionally
independent emission
\begin{align}
y_t \mid z_t \sim F(\theta_{z_t})
\end{align}
for an indexed family of distributions $F(\cdot)$.  Here, $\theta_i$
are the \emph{emission parameters} for mode~$i$. 
\subsubsection*{Switching VAR Processes}
The modeling assumption of the HMM that observations are conditionally independent given the latent mode sequence is often insufficient in capturing the temporal dependencies present in many datasets.  Instead, one can assume that the observations have conditionally \emph{linear} dynamics.  The latent HMM dynamical mode then models switches between a set of such linear dynamical systems in order to capture more complex dynamical phenomena.  We restrict our attention in this paper to switching vector autoregressive (VAR) processes, or \emph{autoregressive HMMs} (AR-HMMs), which are broadly applicable in many domains while maintaining a number of simplifying properties that make them a practical choice computationally.  

We define an AR-HMM, with switches between order-$r$ vector autoregressive processes~\footnote{We denote an order-$r$ VAR process by VAR($r$).}, as
\begin{equation}
\begin{aligned}
\BF{y}_t &= \sum_{i=1}^r A_{i,z_t}\BF{y}_{t-i} + \BF{e}_t(z_t),
\end{aligned}
\label{eqn:SVAR}
\end{equation}
where $z_t$ represents the HMM latent dynamical mode of the system at time $t$, and is defined as in Eq.~\eqref{eqn:HMMmode}.  The mode-specific additive noise term is distributed as $\BF{e}_t(z_t) \sim \mathcal{N}(0,\Sigma_{z_t})$.  We refer to $\BF{A}_k = \{A_{1,k},\dots,A_{r,k}\}$ as the set of \emph{lag matrices}.  Note that the standard HMM with Gaussian emissions arises as a special case of this model when $\BF{A}_{k}=\BF{0}$ for all~$k$.
\subsection{Relating Multiple Time Series}
\label{sec:multipleTimeSeries}
In our applications of interest, we are faced with a \emph{collection} of $N$ time series representing realizations of related dynamical phenomena.  We assume that each time series is individually modeled via a switching VAR process, as in Equation~\eqref{eqn:SVAR}.  Denote the VAR parameters for the $k^{th}$ dynamical mode as $\theta_k = \{\BF{A}_k,\Sigma_k\}$, and assume that we have an unbounded set of possible VAR models $\{\theta_1,\theta_2,\dots\}$.  For example, these parameters might each define a linear motion model for the behaviors \emph{walking}, \emph{running}, \emph{jumping}, and so on; our time series are then each modeled as Markov switches between these behaviors.  We will sometimes avail ourselves the convenient shorthand of referring to $k$ itself as a ``behavior,'' where the intended meaning is the VAR model parameterized by $\theta_k$.  

The way in which our $N$ time series are related is by the overlap in the set of dynamic behaviors that each exhibits.  For example, imagine that our $N$ time series represent observation sequences from the exercise routines of $N$ people.  We expect there to be some overlap in the behaviors exhibited, but also some variability---e.g., some people may solely switch between walking and running, while others switch between running and jumping.  

One can represent the set of behaviors available to each of the time series models with a list of binary \emph{features}.  In particular, let $f_i = [f_{i1}, \, f_{i2}, \ldots]$ denote a binary feature vector for the $i^{th}$ time series.  Setting $f_{ik}=1$ implies that time series $i$ exhibits behavior~$k$ for some subset of values $t \in \{1,\dots,T_i\}$, where $T_i$ is the length of the $i^{th}$ time series.  Our proposed featural model defines $N$ such infinite-dimensional feature vectors, one for each time series.  By discovering the pattern of behavior-sharing via a featural model (i.e., discovering $f_{ik}=f_{jk}=1$ for some $i,j,k$), we can interpret how the time series relate to one another in addition to harnessing the shared structure to pool observations from the same behavior, thus improving our estimate of $\theta_k$.
\subsection{Beta Processes}
\label{background:BetaProcess}
Inferring the structure of behavior sharing within a Bayesian framework requires defining a prior on the feature inclusion probabilities.  Since we want to maintain an unbounded set of possible behaviors (and thus require infinite-dimensional feature vectors), we appeal to a Bayesian nonparametric featural model based on the \emph{beta process-Bernoulli process}.  Informally, one can think of the beta process as defining an infinite set of coin-flipping probabilities and each Bernoulli process realization is the outcome from an infinite coin-flipping sequence based on the beta-process-determined coin weights.  The set of resulting \emph{heads} indicate the set of selected \emph{features}, and implicitly defines an infinite-dimensional \emph{feature vector}.  The properties of the beta process induce sparsity in the feature space by encouraging sharing of features among the Bernoulli process realizations.  The inherent conjugacy of the beta process to the Bernoulli process allows for an analytic predictive distribution on a feature vector (i.e., Bernoulli realization) based on the feature vectors observed so far (i.e., previous Bernoulli process draws).  As outlined in Section~\ref{sec:IBP}, this predictive distribution can be described via the Indian buffet process under certain parameterizations of the beta process.

\subsubsection*{The Beta Process - Bernoulli Process Featural Model}

\begin{figure}[t]
  \centering
\begin{tabular}{cc}
  \includegraphics[width=0.5\columnwidth]{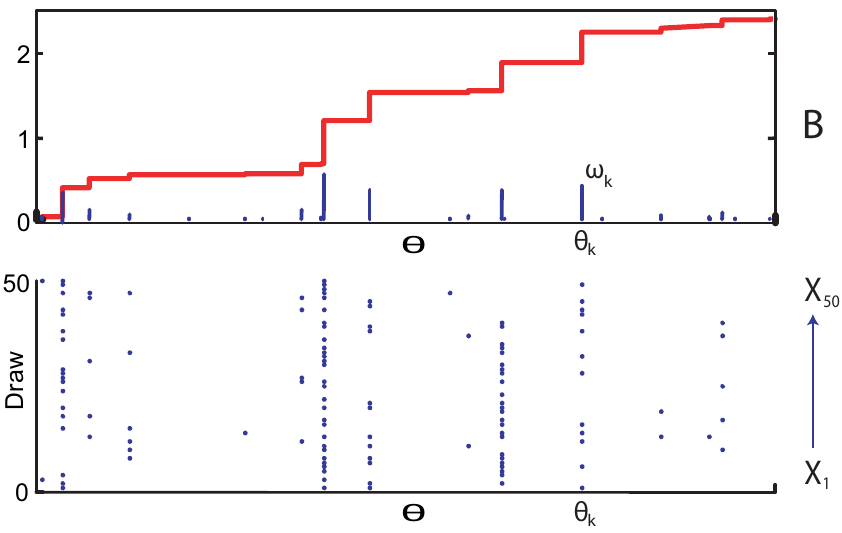}
& \includegraphics[width =
0.4\columnwidth]{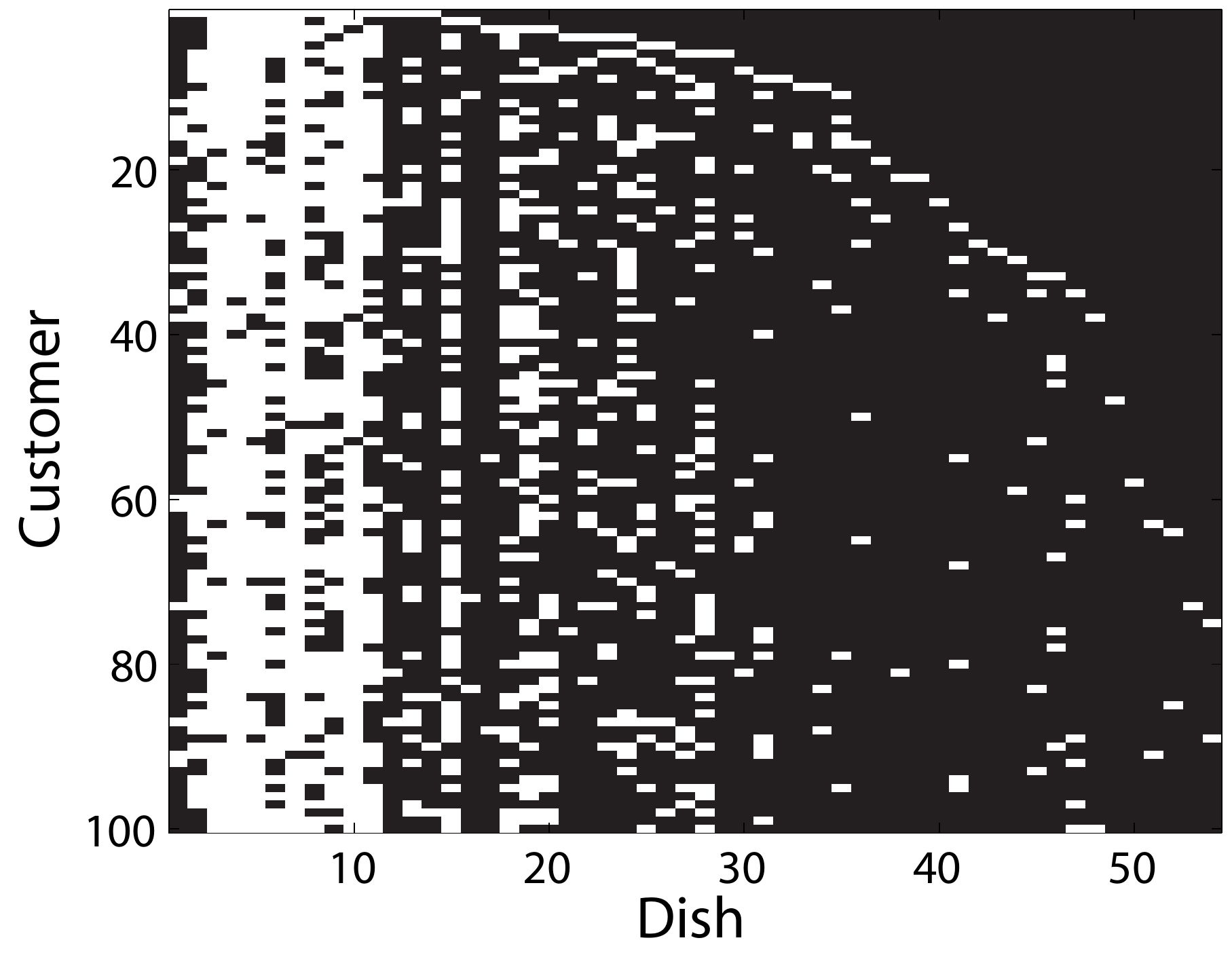}\\
(a) & (b)
\end{tabular}
  \caption[A draw from a beta process, and associated
Bernoulli realizations, along with a realization from the Indian
buffet process.]{(a) \textit{Top}: A draw $B$ from a beta process is
shown in blue, with the corresponding cumulative distribution in
red. \textit{Bottom}: 50 draws $X_i$ from a Bernoulli process using
the beta process realization. Each blue dot corresponds to a
coin-flip at that atom in $B$ that came up heads. (b) An image of a
feature matrix associated with a realization from an Indian buffet
process with $\alpha=10$. Each row corresponding to a different
customer, and each column a different dish.  White indicates a
chosen feature.} \label{fig:BPBePrealizations}
\end{figure}

The beta process is a special case of a general class of stochastic processes known as~\emph{completely random measures}~\citep{Kin1967}. A completely random measure $B$ is defined such that for any disjoint sets $A_1$ and $A_2$ (of some sigma algebra $\mathcal{A}$ on a measurable space $\Theta$), the corresponding random measures $B(A_1)$ and $B(A_2)$ are independent. This idea generalizes the family of \emph{independent increments processes} on the real line. All completely random measures can be constructed from realizations of a nonhomogenous Poisson process (up to a deterministic component)~\citep{Kin1967}. Specifically, a Poisson rate measure $\eta$ is defined on a product space $\Theta \otimes \mathbb{R}$, and a draw from the specified Poisson process yields a collection of points $\{\theta_j,\omega_j\}$ that can be used to define a completely random measure:
\begin{align}
	B = \sum_{k=1}^\infty \omega_k\delta_{\theta_k}.
	\label{eqn:CRM}
\end{align}
This construction assumes $\eta$ has infinite mass, yielding the countably infinite collection of points from the Poisson process.  From Eq.~\eqref{eqn:CRM}, we see that completely random measures are discrete. Consider a rate measure defined as the product of an arbitrary sigma-finite \emph{base measure} $B_0$, with total mass $B_0(\Theta)=\alpha$, and an improper beta distribution on the product space $\Theta \otimes [0,1]$:
\begin{equation}
	\nu(d\omega, d\theta) = c\omega^{-1}(1 - \omega)^{c-1}d\omega B_0(d\theta),
\end{equation}
where $c>0$ is referred to as a \emph{concentration parameter}. The resulting completely random measure is known as the \emph{beta process} with draws denoted by $B \sim \mbox{BP}(c,B_0)$~\footnote{Letting the rate measure be defined as a product of a base measure $G_0$ and an improper gamma distribution $\eta(d\theta,d\omega) = cp^{-1}e^{-cp}dpG_0(d\theta)$, with $c>0$, gives rise to completely random measures $G\sim \mbox{GP}(c,G_0)$, where GP denotes a \emph{gamma process}. Normalizing $G$ yields draws from a Dirichlet process $\DP{\alpha}{G_0/\alpha}$, with $\alpha = G_0(\Theta)$. Note that these random \emph{probability} measures $G$ are necessarily not completely random since the random variables $G(A_1)$ and $G(A_2)$ for disjoint sets $A_1$ and $A_2$ are dependent due to the normalization constraint.}. Note that using this construction, the weights $\omega_k$ of the atoms in $B$ lie in the interval $(0,1)$. Since $\eta$ is $\sigma$-finite, Campbell's theorem~\citep{Kingman:93} guarantees that for $\alpha$ finite, $B$ has finite expected measure. For an example realization and its associated cumulative distribution, see Fig.~\ref{fig:BPBePrealizations}.
%

Note that for a base measure $B_0$ containing atoms, a sample $B \sim \mbox{BP}(c,B_0)$ necessarily contains each of these atoms $\theta_k$ with associated weights
\begin{align}
\omega_k \sim \mbox{Beta}(c q_k, c(1-q_k)),
\end{align}
where $q_k \in (0,1)$ denotes the mass of the $k^{th}$ atom in $B_0$. 

The beta process is conjugate to a class of \emph{Bernoulli processes}~\citep{Thibaux:07}, denoted by $\mbox{BeP}(B)$, which provide our sought-for featural representation. A realization
\begin{align}
X_i\mid B \sim \mbox{BeP}(B),
\end{align}
with $B$ an atomic measure, is a collection of unit-mass atoms on $\Theta$ located at some subset of the atoms in $B$. In particular,
\begin{align}
f_{ik} \sim \mbox{Bernoulli}(\omega_k)
\end{align}
is sampled independently for each atom $\theta_k$ in $B$~\footnote{One can visualize this process as walking along the atoms of a discrete measure $B$ and, at each atom $\theta_k$, flipping a coin with probability of heads given by $\omega_k$.}, and then
\begin{align}
X_i = \sum_k f_{ik} \delta_{\theta_k}.
\end{align}
Example realizations of $X_i \sim \mbox{BeP}(B)$, with $B$ a draw from a beta process, are shown in Fig.~\ref{fig:BPBePrealizations}(a).

For continuous measures $B$, we draw $L \sim \mbox{Poisson}(B(\Theta))$ and then independently sample a set of $L$ atoms $\theta_\ell \sim B(\Theta)^{-1}B$. The Bernoulli realization is then given by:
\begin{align}
X_i = \sum_{\ell=1}^L \delta_{\theta_\ell}.
\end{align}

In our subsequent development, we interpret the atom locations $\theta_k$ as a set of global features that can be shared among multiple time series. A Bernoulli process realization $X_i$ then determines the subset of features allocated to time series~$i$:
\begin{align}
	B \mid B_0, c&\sim \mbox{BP}(c,B_0)\nonumber\\
	X_i \mid B &\sim \mbox{BeP}(B), \quad i=1,\dots,N. \label{eqn:HierarchicalBeta} 
\end{align}

Computationally, Bernoulli process realizations $X_i$ are often summarized by an infinite vector of binary indicator variables $\fset{i} = [f_{i1}, f_{i2}, \ldots]$, where $f_{ik}=1$ if and only if time series~$i$ exhibits feature~$k$. Using the beta process measure $B$ to tie together the feature vectors encourages them to share similar features while allowing time-series-specific variability.
\section{Describing Multiple Time Series with Beta Processes} \label{sec:model}
We employ the beta process featural model of Section~\ref{background:BetaProcess} to define a prior on the collection of \emph{infinite}-dimensional feature vectors $\fset{i} = [f_{i1}, \, f_{i2}, \ldots]$ used to describe the relationship amongst our $N$ time series.  Recall from Section~\ref{sec:multipleTimeSeries} that the globally-shared parameters $\theta_k$ define the possible \emph{behaviors} (e.g., VAR processes), while the feature vector $\fset{i}$ indicates the behaviors exhibited by time series $i$.    

\subsubsection*{Beta Process Prior on Features}
In our scenario, the beta process hierarchy of Equation~\eqref{eqn:HierarchicalBeta} can be interpreted as follows.  The random measure $B \sim \mbox{BP}(c,B_0)$ defines a set of weights on the global collection of behaviors.  Then, each \emph{time series} $i$ is associated with a draw from a Bernoulli process, $X_i\mid B \sim \mbox{BeP}(B)$.  The Bernoulli process realization $X_i = \sum_k f_{ik} \delta_{\theta_k}$ implicitly defines the feature vector $\fset{i}$ for time series $i$, indicating which set of globally-shared behaviors that time series has selected.  Such a featural model seeks to allow for infinitely many possible behaviors, while encouraging a sparse, finite representation and flexible sharing of behaviors between time series.  For example, the lower subfigure in Fig.~\ref{fig:BPBePrealizations}(a) illustrates a collection of feature vectors drawn from this process.

Conditioned on the set of $N$ feature vectors $\fset{i}, i=1,\dots,N$ drawn from the hierarchy of Equation~\eqref{eqn:HierarchicalBeta}, the model reduces to a collection of $N$ switching VAR processes, each defined on the finite state space formed by the set of selected behaviors for that time series.  In the following section, we define the generative process for the Markov dynamics based on the sampled feature vectors.

\subsubsection*{Feature-Constrained Transition Distributions}
Given $\fset{i}$, the $i^{th}$ time series's Markov transitions among its set of dynamic behaviors are governed by a set of \emph{feature-constrained transition distributions} \mbox{$\BF{\pi}^{(i)} =\{\symsubsup{\pi}{k}{i}\}$}. In particular, motivated by the fact that Dirichlet-distributed probability mass functions can be generated via normalized gamma random variables, for each time series $i$ we define a doubly infinite collection of random variables:
\begin{equation}
	\symsubsup{\eta}{jk}{i}\mid \gamma,\kappa \sim \mbox{Gamma}(\gamma+\kappa\delta(j,k),1), \label{eqn:transitionGamma} 
\end{equation}
Here, $\delta(j,k)$ indicates the Kronecker delta function. Using this collection of \emph{transition variables}, denoted by $\BF{\eta}^{(i)}$, one can define time-series-specific, feature-constrained transition distributions:
\begin{align}
	\symsubsup{\pi}{j}{i} = \frac{
	\begin{bmatrix}
		\symsubsup{\eta}{j1}{i} & \symsubsup{\eta}{j2}{i} & \dots\; 
	\end{bmatrix}
	\otimes \fset{i}}{\sum_{k|f_{ik}=1} \symsubsup{\eta}{jk}{i}}, \label{eqn:normEta} 
\end{align}
where $\otimes$ denotes the element-wise, or Hadamard, vector product.  This construction defines $\symsubsup{\pi}{j}{i}$ over the full set of positive integers, but assigns positive mass only at indices~$k$ where $f_{ik}=1$, thus constraining time series $i$ to solely transition amongst the dynamical behaviors indicated by its feature vector $\fset{i}$.

The preceding generative process can be equivalently represented via a sample $\symsubsup{\tilde{\pi}}{j}{i}$ from a finite Dirichlet distribution of dimension $K_i = \sum_k f_{ik}$, containing the non-zero entries of $\symsubsup{\pi}{j}{i}$:
\begin{equation}
	\symsubsup{\tilde{\pi}}{j}{i} \mid \fset{i}, \gamma,\kappa \sim \mbox{Dir}([\gamma,\dots,\gamma,\gamma + \kappa,\gamma,\dots\gamma]). \label{eqn:DirPrior} 
\end{equation}
The $\kappa$ hyperparameter places extra expected mass on the component of $\symsubsup{\tilde{\pi}}{j}{i}$ corresponding to a self-transition $\symsubsup{\pi}{jj}{i}$, analogously to the sticky hyperparameter of the sticky HDP-HMM~\citep{Fox:AOAS11}. We also use the representation
\begin{align}
	\symsubsup{\pi}{j}{i} \mid \fset{i},\gamma,\kappa \sim \mbox{Dir}([\gamma,\dots,\gamma,\gamma + \kappa,\gamma,\dots]\otimes \fset{i}),\label{eqn:DirPrior2} 
\end{align}
implying $\symsubsup{\pi}{j}{i} = 
\begin{bmatrix}
	\symsubsup{\pi}{j1}{i} & \symsubsup{\pi}{j2}{i} & \dots 
\end{bmatrix}$, with only a finite number of non-zero entries $\symsubsup{\pi}{jk}{i}$. This representation is really an abuse of notation since the Dirichlet distribution is not defined for infinitely many parameters. In reality, we are simply examining a $K_i$-dimensional Dirichlet distribution as in Eq.~\eqref{eqn:DirPrior}. However, the notation of Eq.~\eqref{eqn:DirPrior2} is useful in reminding the reader that the indices of $\symsubsup{\tilde{\pi}}{j}{i}$ defined by Eq.~\eqref{eqn:DirPrior} are not over 1 to $K_i$, but rather over the $K_i$ values of $k$ such that $f_{ik}=1$. Additionally, this notation is useful for concise representations of the posterior distribution.

\subsubsection*{VAR Likelihoods}
Although the methodology described thus far applies equally well to HMMs and other Markov switching processes, henceforth we focus our attention on the AR-HMM and develop the full model specification and inference procedures needed to treat our motivating example of visual motion capture.  Specifically, let $\symsubsupB{y}{t}{i}$ represent the observed value of the $i^{th}$ time series at time $t$, and let $\symsubsup{z}{t}{i}$ denote the latent dynamical mode. Assuming an order-$r$ AR-HMM, we have
\begin{equation}
	\begin{aligned}
		\symsubsup{z}{t}{i} &\sim \symsubsup{\pi}{\symsubsup{z}{t-1}{i}}{i}\\
		\symsubsupB{y}{t}{i} &= \sum_{j=1}^r A_{j,\symsubsup{z}{t}{i}}\symsubsupB{y}{t-j}{i} + \symsubsupB{e}{t}{i}(\symsubsup{z}{t}{i}) \triangleq \BF{A}_{\symsubsup{z}{t}{i}}\symsubsupB{\tilde{y}}{t}{i} + \symsubsupB{e}{t}{i}(\symsubsup{z}{t}{i}), 
	\end{aligned}
	\label{eqn:multSVAR} 
\end{equation}
where $\symsubsupB{e}{t}{i}(k) \sim \mathcal{N}(0,\Sigma_{k})$, $\BF{A}_{k} = 
\begin{bmatrix}
	A_{1,k} & \dots & A_{r,k}
\end{bmatrix}
$, and $\symsubsupB{\tilde{y}}{t}{i} = 
\begin{bmatrix}
	\smash{\symsubsupT{y}{t-1}{i}} & \dots & \smash{\symsubsupT{y}{t-r}{i}}
\end{bmatrix}^T$. Recall that each of the behaviors $\theta_k = \{\BF{A}_{k},\Sigma_k\}$ defines a different VAR($r$) dynamical mode and the feature-constrained transition distributions $\pi^{(i)}$ restrict time series $i$ to select among dynamic behaviors (indexed at time $t$ by $\symsubsup{z}{t}{i}$) that were picked out by its feature vector $\fset{i}$. Our beta-process-based featural model couples the dynamic behaviors exhibited by different time series. 

\subsubsection*{Prior on VAR Parameters}
To complete the Bayesian model specification, a conjugate matrix-normal inverse-Wishart (MNIW) prior (cf.,~\cite{West}) is placed on the shared collection of dynamic parameters $\theta_k = \{\BF{A}_k,\Sigma_k\}$.  Specifically, this prior is comprised of an inverse Wishart prior on $\Sigma_k$ and (conditionally) a matrix normal prior on $\BF{A}_k$:
\begin{equation}
\begin{aligned}
 \Sigma_k \mid n_0,S_0 &\sim \mbox{IW}(n_0,S_0)\\
 \BF{A}_k \mid \Sigma_k,M,K &\sim \MN{\BF{A}_k}{M}{\Sigma_k}{K},
\end{aligned}
\end{equation} 
with $n_0$ the degrees of freedom, $S_0$ the scale matrix, $M$ the mean dynamic matrix, and $K$ a covariance matrix that together with $\Sigma_k$ defines the covariance of $A_k$.  This prior defines the base measure $B_0$ up to the total mass parameter $\alpha$, which has to be separately assigned.  As motivated in Section~\ref{sec:IBPhyperparameters}, this latter parameter is given a gamma prior.

Since the library of possible dynamic parameters is shared by all time series, posterior inference of each parameter set $\theta_k$ relies on pooling data amongst the time series that have $f_{ik}=1$. It is through this pooling of data that one may achieve more robust parameter estimates than from considering each time series individually.  

\subsubsection*{The BP-AR-HMM}
\begin{figure}
	[t!] \centering \hspace{0.2in} 
	\includegraphics[height=2.5in]{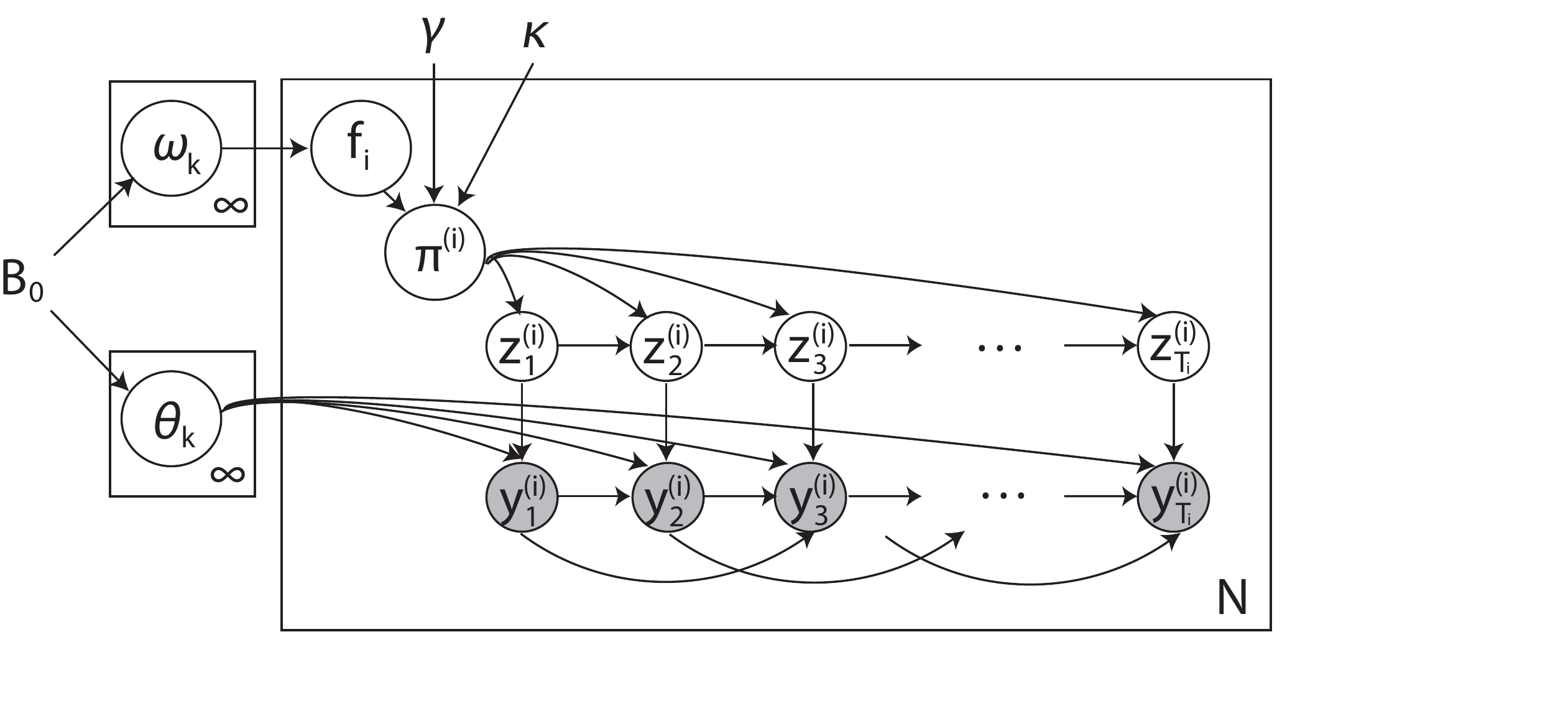}\vspace{-0.2in} \caption[Graphical model of the BP-AR-HMM.] {Graphical model of the BP-AR-HMM. The beta process distributed measure $\mbox{$B \mid B_0 \sim \mbox{BP}(1,B_0)$}$ is represented by its masses $\omega_k$ and locations $\theta_k$, as in Eq.~\eqref{eqn:CRM}. The features are then conditionally independent draws $\mbox{$f_{ik} \mid \omega_k \sim \mbox{Bernoulli}(\omega_k)$}$, and are used to define feature-constrained transition distributions $\mbox{$\symsubsup{\pi}{j}{i} \mid \fset{i}, \gamma,\kappa \sim \mbox{Dir}([\gamma,\dots,\gamma,\gamma+\kappa,\gamma,\dots]\otimes \fset{i})$}$. The switching VAR dynamics are as in Eq.~\eqref{eqn:multSVAR}.} \label{fig:BPARHMM} 
\end{figure}

We term the resulting model the \emph{BP-autoregressive-HMM} (BP-AR-HMM), with a graphical model representation presented in Fig.~\ref{fig:BPARHMM}.  Considering the \emph{feature space} (i.e., set of autoregressive parameters) and the \emph{temporal dynamics} (i.e., set of transition distributions) as separate dimensions, one can think of the BP-AR-HMM as a spatio-temporal process comprised of a (continuous) beta process in space and discrete-time Markovian dynamics in time.  The overall model specification is summarized as:
\begin{equation}
	\begin{aligned}
		B \mid B_0 &\sim \mbox{BP}(1,B_0)\\
		X_i \mid B &\sim \mbox{BeP}(B), \quad i = 1,\dots, N\\
		\symsubsup{\pi}{j}{i} \mid \fset{i},\gamma,\kappa &\sim \mbox{Dir}([\gamma,\dots,\gamma,\gamma + \kappa,\gamma,\dots]\otimes \fset{i}), \quad i=1,\dots,N, \,\, j=1,2,\dots\\
		\symsubsup{z}{t}{i} &\sim \symsubsup{\pi}{\symsubsup{z}{t-1}{i}}{i}, \quad i=1,\dots,N, \,\, t=1,\dots,T_i\\
		\symsubsupB{y}{t}{i} &= \BF{A}_{\symsubsup{z}{t}{i}}\symsubsupB{\tilde{y}}{t}{i} + \symsubsupB{e}{t}{i}(\symsubsup{z}{t}{i}),  \quad i=1,\dots,N, \,\, t=1,\dots,T_i.
	\end{aligned}
	\label{eqn:BPARHMM} 
\end{equation}
\section{MCMC Posterior Computations} \label{sec:MCMC}
In this section, we develop an MCMC method which alternates between sampling binary feature assignments given observations and dynamic parameters, and sampling dynamic parameters given observations and features. The sampler interleaves Metropolis-Hastings and Gibbs sampling updates, which are sometimes simplified by appropriate auxiliary variables. We leverage the fact that fixed feature assignments instantiate a set of \emph{finite} AR-HMMs, for which dynamic programming can be used to efficiently compute marginal likelihoods. Computationally, sampling the potentially infinite set of time-series-specific features in our beta process featural model relies on a predictive distribution on features that can be described via the \emph{Indian buffet process} (IBP)~\citep{GriffithsGhahramani:05}.  The details of the IBP are outlined below.  As a key component of our feature-sampling, we introduce a new approach employing incremental ``birth'' and ``death'' proposals, improving on previous exact samplers for IBP models in the non-conjugate case~\citep{Meeds:07}.
\subsection{Background: The Indian Buffet Process}
\label{sec:IBP}
As shown by~\citet{Thibaux:07}, marginalizing over the latent beta process $B$ in the hierarchical model of Equation~\eqref{eqn:HierarchicalBeta}, and taking $c=1$, induces a predictive distribution on feature indicators known as the Indian buffet process (IBP)~\citep{GriffithsGhahramani:05}. The IBP is based on a culinary metaphor in which customers arrive at an infinitely long buffet line of dishes, or features (\emph{behaviors} in our case). The first arriving customer, or \emph{time series} in our case, chooses $\mbox{Poisson}(\alpha)$ dishes. Each subsequent customer~$i$ selects a previously tasted dish~$k$ with probability $m_k/i$ proportional to the number of previous customers $m_k$ to sample it, and also samples $\mbox{Poisson}(\alpha/i)$ new dishes. The feature matrix associated with a realization from an Indian buffet process is shown in Fig.~\ref{fig:BPBePrealizations}(b).

To derive the IBP from the beta process formulation described above,
we note that the probability $X_{i}$ contains feature $\theta_k$
after having observed $X_1,\dots,X_{i-1}$ is equal to the expected
mass of that atom:
\begin{align}
p(f_{ik}=1 \mid X_1,\dots,X_{i-1}) = \mathbb{E}_{B\mid
X_1,\dots,X_{i-1}}[ p(f_{ik}=1\mid B)] = \mathbb{E}_{B\mid
X_1,\dots,X_{i-1}}[\omega_k],
\end{align}
where our notation $\mathbb{E}_{B}[\cdot]$ means to take the
expectation with respect to the distribution of $B$. Because beta process priors are conjugate to the Bernoulli process~\citep{Kim:99}, the posterior distribution given $N$ samples $X_i \sim \mbox{BeP}(B)$ is a beta process with updated parameters:
\begin{align}
B \mid X_1,\dots,X_N, B_0, c &\sim
\mbox{BP}\Bigg(c+N,\frac{c}{c+N}B_0 + \frac{1}{c+N}\sum_{i=1}^N
X_i\Bigg)\\
&= \mbox{BP}\Bigg(c+N,\frac{c}{c+N}B_0 + \sum_{k=1}^{\Kplus}
\frac{m_k}{c+N}\delta_{\theta_k}\Bigg). \label{eqn:betapost}
\end{align}
Here, $m_k$ denotes the number of time series $X_i$ that select the $k^{th}$ feature $\theta_k$ (i.e., $f_{ik}=1$).  For simplicity, we have reordered the feature indices to list first the $\Kplus$ features used by at least one time series.

Using the
posterior distribution defined in Eq.~\eqref{eqn:betapost}, we
consider the discrete and continuous portions of the base measure
separately.  The discrete component is a collection of atoms at
locations $\theta_1,\dots,\theta_{K_+}$, each with weight
\begin{align}
q_k = \frac{m_k}{c+i-1},
\end{align}
where $K_+$ is the number of unique atoms present in
$X_1,\dots,X_{i-1}$. For each of the currently instantiated features
$k \in \{1,\dots,K_+\}$, we have
\begin{align}
\omega_k \sim \mbox{Beta}((c+i-1)q_k,(c+i-1)(1-q_k))
\end{align}
such that the expected weight is simply $q_k$, implying that the
$i^{th}$ time series chooses one of the currently instantiated features
with probability proportional to the number of time series that already
chose that feature, $m_k$. We now consider the continuous portion of
the base measure,
\begin{align}
\frac{c}{c+i-1}B_0.
\end{align}
The Poisson process defined by this rate function generates
\begin{align}
\mbox{Poisson}\left(\frac{c}{c+i-1}B_0(\Theta)\right) =
\mbox{Poisson}\left(\frac{c}{c+i-1}\alpha\right)
\end{align}
new atoms in $X_i$ that do not appear in $X_1,\dots,X_{i-1}$.
Following this argument, the first time series simply chooses
$\mbox{Poisson}(\alpha)$ features. If we specialize this process to
$c=1$, we arrive at the IBP.
\subsection{Sampling binary feature assignments} \label{sec:featureSampling}
Let $\BF{F}^{-ik}$ denote the set of all binary feature indicators excluding $f_{ik}$, and $K_+^{-i}$
be the number of behaviors used by all of the other time series~\footnote{Some of the $K_+^{-i}$ features may also be used by time series $i$, but only those not unique to that time series.}. For notational simplicity, we assume that these behaviors are indexed by $\{1,\dots,K_+^{-i}\}$. The IBP prior differentiates between features, or behaviors, that other time series have already selected and those unique to the current time series. Thus, we examine each of these cases separately.

\subsubsection*{Shared features} Given the $i^{th}$ time series $\symsubsupB{y}{1:T_i}{i}$, transition variables $\BF{\eta}^{(i)} = \symsubsup{\eta}{1:K_+^{-i},1:K_+^{-i}}{i}$, and shared dynamic parameters $\theta_{1:K_+^{-i}}$, the feature indicators $f_{ik}$ for currently used features $k \in \{1,\dots,K_+^{-i}\}$ have the following posterior distribution:
\begin{equation}
	p(f_{ik}\mid \BF{F}^{-ik}\!,\symsubsupB{y}{1:T_i}{i},\BF{\eta}^{(i)}\!, \theta_{1:K_+^{-i}},\alpha) \propto p(f_{ik}\mid \BF{F}^{-ik}\!, \alpha) p(\symsubsupB{y}{1:T_i}{i}\mid \fset{i}, \BF{\eta}^{(i)}\!,\theta_{1:K_+^{-i}}). \label{eqn:Fsampling} 
\end{equation}
Here, the IBP prior described in Section~\ref{background:BetaProcess} implies that $p(f_{ik}=1\mid \BF{F}^{-ik}\!, \alpha) = m_k^{-i}/N$, where $m_k^{-i}$ denotes the number of time series \emph{other} than time series $i$ that exhibit behavior $k$. In evaluating this expression, we have exploited the exchangeability of the IBP~\citep{GriffithsGhahramani:05}, which follows directly from the beta process construction~\citep{Thibaux:07}.
For binary random variables, Metropolis-Hastings proposals can mix faster~\citep{Frigessi:93} and have greater efficiency~\citep{Liu:96} than standard Gibbs samplers. To update $f_{ik}$ given $\BF{F}^{-ik}\!$, we thus use the posterior of Eq.~\eqref{eqn:Fsampling} to evaluate a Metropolis-Hastings proposal which flips $f_{ik}$ to the complement $\bar{f}$ of its current value $f$:
\begin{align}
	f_{ik} &\sim \rho(\bar{f} \mid f)\delta(f_{ik},\bar{f}) + (1-\rho(\bar{f} \mid f))\delta(f_{ik},f) \nonumber \\
	\rho(\bar{f} \mid f) &= \min \Bigg\{\frac{p(f_{ik}=\bar{f}\mid \BF{F}^{-ik}\!,\symsubsupB{y}{1:T_i}{i},\BF{\eta}^{(i)}\!,\theta_{1:K_+^{-i}},\alpha)}{p(f_{ik}=f\mid \BF{F}^{-ik}\!,\symsubsupB{y}{1:T_i}{i},\BF{\eta}^{(i)}\!,\theta_{1:K_+^{-i}},\alpha)},1\Bigg\}. \label{eqn:sharedFeaturesMH} 
\end{align}
To compute likelihoods, we combine $\fset{i}$ and $\BF{\eta}^{(i)}$ to construct feature-constrained transition distributions $\symsubsup{\pi}{j}{i}$ as in Eq.~\eqref{eqn:normEta}, and marginalize over the exponentially large set of possible latent mode sequences by applying a variant of the sum-product message passing algorithm for AR-HMMs.  (See Appendix~\ref{app:sumprod}.) 

\subsubsection*{Unique features} An alternative approach is needed to sample the $\mbox{Poisson}(\alpha/N)$ ``unique'' features associated only with time series~$i$. Let $K_+ = K_+^{-i}+
\newfeatures{i}$, where $
\newfeatures{i}$ is the number of unique features chosen, and define $\fset{-i} = f_{i,1:K_+^{-i}}$ and $\fset{+i} = f_{i,K_+^{-i}+1:K_+}$. The posterior distribution over $
\newfeatures{i}$ is then given by
\vspace{-0.1in} 
\begin{multline}
	p(
	\newfeatures{i} \mid \fset{i},\symsubsupB{y}{1:T_i}{i},\BF{\eta}^{(i)}\!,\theta_{1:K_+^{-i}},\alpha) \propto \frac{(\frac{\alpha}{N})^{
	\newfeatures{i}}e^{-\frac{\alpha}{N}}}{
	\newfeatures{i}!}\\
	\iint p(\symsubsupB{y}{1:T_i}{i}\mid \fset{-i},\fset{+i}=\BF{1}, \BF{\eta}^{(i)}\!,\BF{\eta}_+,\theta_{1:K_+^{-i}},\BF{\theta}_+) \;dB_0(\BF{\theta}_+)dH(\BF{\eta}_+), 
\end{multline}
where $H$ is the gamma prior on transition variables $\symsubsup{\eta}{jk}{i}$, and we recall that $B_0$ is the base measure of the beta process. The set $\BF{\theta}_+ = \theta_{K_+^{-i} + 1: K_+}$ consists of the parameters of unique features, and $\BF{\eta}_+$ the transition parameters $\symsubsup{\eta}{jk}{i}$ to or from unique features $j,k \in \{K_+^{-i} + 1:K_+\}$. Exact evaluation of this integral is intractable due to dependencies induced by the AR-HMMs.

One early approach to approximate Gibbs sampling in non-conjugate IBP models relies on a finite truncation of the limiting Bernoulli process~\citep{Gorur:06}. That is, drawing $n_i \sim \mbox{Poisson}(\alpha/N)$ distribution is equivalent to setting $n_i$ equal to the number of successes in infinitely many Bernoulli trials, each with probability of success
\begin{align}
	\lim_{K \rightarrow \infty} \frac{\alpha/K}{\alpha/K + N}. 
\end{align}
\citet{Gorur:06} truncate this process and instead consider $K^*$ Bernoulli trials with probability $(\alpha/K^*)/(\alpha/K^* + N)$. \citet{Meeds:07} instead consider independent Metropolis proposals which replace the existing unique features by $n_i \sim \mbox{Poisson}(\alpha/N)$ new features, with corresponding parameters $\BF{\theta}_+$ drawn from the prior. For high-dimensional models such as those considered in this paper, however, such moves have extremely low acceptance rates.

We instead develop a birth and death reversible jump MCMC sampler~\citep{Green:95}, which proposes to either add a single new feature, or eliminate one of the existing features in $\fset{+i}$. Our proposal distribution factors as follows:
\begin{equation}
	q(\fset{+i}',\BF{\theta}_{+}',\BF{\eta}_{+}' \mid \fset{+i},\BF{\theta}_{+},\BF{\eta}_{+}) = q_f(\fset{+i}' \mid \fset{+i}) q_{\theta}(\BF{\theta}_{+}' \mid \fset{+i}',\fset{+i},\BF{\theta}_{+}) q_{\eta}(\BF{\eta}_{+}' \mid \fset{+i}', \fset{+i}, \BF{\eta}_{+})\label{eqn:jointproposal} 
\end{equation}
Let $n_i = \sum_k\! f_{+ik}$. The feature proposal $q_f(\cdot\mid\cdot)$ encodes the probabilities of birth and death moves, which we set as follows: A new feature is created with probability $0.5$, and each of the $n_i$ existing features is deleted with probability $0.5/n_i$. This set of possible proposals leads to considering transitions from $n_i$ to $n_i'$ unique features, with $n_i'=n_i+1$ in the case of a birth proposal, or $n_i'=n_i-1$ in the case of a proposed feature death. Note that if the proposal from the distribution defined in Eq.~\eqref{eqn:jointproposal} is rejected, we maintain $n_i' = n_i$ unique features. For parameters, we define our proposal using the generative model:
\begin{align}
	q_{\theta}(\BF{\theta}_{+}' \mid \fset{+i}',\fset{+i},\BF{\theta}_{+}) = \left\{ 
	\begin{array}{ll}
		b_0(\theta_{+,n_i+1}') \prod_{k=1}^{n_i} \delta_{\theta_{+,k}}(\theta_{+,k}'), & \hbox{birth of feature } n_i + 1; \\
		\prod_{k\neq \ell}\delta_{\theta_{+,k}}(\theta_{+,k}'), & \hbox{death of feature } \ell. 
	\end{array}
	\right.\label{eqn:parameter_proposals} 
\end{align}
That is, for a birth proposal, a new parameter $\theta_{+,n_i+1}'$ is drawn from the prior and all other parameters remain the same. For a death proposal of feature $j$, we simply eliminate that parameter from the model. Here, $b_0$ is the density associated with $\alpha^{-1}B_0$. The distribution $q_{\eta}(\cdot\mid\cdot)$ is defined similarly, but using the gamma prior on transition variables of Eq.~\eqref{eqn:transitionGamma}.

The Metropolis-Hastings acceptance probability is then given by
\begin{equation}
	\rho(\fset{+i}',\BF{\theta}_{+}',\BF{\eta}_{+}' \mid \fset{+i},\BF{\theta}_{+},\BF{\eta}_{+}) = \min\{r(\fset{+i}',\BF{\theta}_{+}',\BF{\eta}_{+}' \mid \fset{+i},\BF{\theta}_{+},\BF{\eta}_{+}),1\}. 
\end{equation}
As derived in Appendix~\ref{app:birthdeath}, we compactly represent the acceptance ratio $r(\cdot \mid \cdot)$ for either a birth or death move as
\begin{equation}
	\frac{p(\symsubsupB{y}{1:T_i}{i}\mid [\fset{-i} \, \fset{+i}'], \theta_{1:K_+},\BF{\theta}_{+}',\BF{\eta}^{(i)}, \BF{\eta}_{+}') \; \mbox{Poisson}(n_i' \mid \alpha/N) \; q_f(\fset{+i} \mid \fset{+i}')}{ p(\symsubsupB{y}{1:T_i}{i}\mid [\fset{-i} \, \fset{+i}], \theta_{1:K_+},\BF{\eta}^{(i)}) \; \mbox{Poisson}(n_i \mid \alpha/N) \; q_f(\fset{+i}' \mid \fset{+i})},\label{eqn:uniqueFeaturesMH} 
\end{equation}
where we recall that $n_i' = \sum_k\! f_{+ik}'$. Because our birth and death proposals do not modify the values of existing parameters, the Jacobian term normally arising in reversible jump MCMC algorithms simply equals one.
\subsection{Sampling dynamic parameters and transition variables}
Posterior updates to transition variables $\BF{\eta}^{(i)}$ and shared dynamic parameters $\theta_k$ are greatly simplified if we instantiate the mode sequences $\symsubsup{z}{1:T_i}{i}$ for each time series $i$. We treat these mode sequences as auxiliary variables that are discarded for subsequent updates of feature assignments $\fset{i}$.

\subsubsection*{Mode sequences $\symsubsup{z}{1:T_i}{i}$} Given feature-constrained transition distributions $\BF{\pi}^{(i)}$ and dynamic parameters $\{\theta_k\}$, along with the observation sequence $\symsubsupB{y}{1:T_i}{i}$, we block sample the mode sequence $\symsubsup{z}{1:T_i}{i}$ by computing backward messages $m_{t+1,t}(\symsubsup{z}{t}{i}) \propto p(\symsubsupB{y}{t+1:T_i}{i} \mid \symsubsup{z}{t}{i},\symsubsupB{\tilde{y}}{t}{i},\BF{\pi}^{(i)},\{\theta_k\})$, and then recursively sampling each $\symsubsup{z}{t}{i}$:
\begin{equation}
	\symsubsup{z}{t}{i} \mid \symsubsup{z}{t-1}{i}, \symsubsupB{y}{1:T_i}{i},\BF{\pi}^{(i)}\!, \{\theta_k\} \sim \symsubsup{\pi}{\symsubsup{z}{t-1}{i}}{i}\!(\symsubsup{z}{t}{i}) \mathcal{N}\big(\symsubsupB{y}{t}{i}; \BF{A}_{\symsubsup{z}{t}{i}}\symsubsupB{\tilde{y}}{t}{i}, \Sigma_{\symsubsup{z}{t}{i}}\big) m_{t+1,t}(\symsubsup{z}{t}{i}). 
\end{equation}
This backward message-passing, forward-sampling scheme is detailed in Appendix~\ref{app:sumprod}.

\subsubsection*{Transition distributions $\symsubsup{\pi}{j}{i}$}
We use the fact that Dirichlet priors are conjugate to multinomial observations $\symsubsup{z}{1:T}{i}$ to derive the posterior of $\symsubsup{\pi}{j}{i}$ as
\begin{align}
	\symsubsup{\pi}{j}{i} \mid \fset{i},\symsubsup{z}{1:T}{i}, \gamma,\kappa \sim \mbox{Dir}([\gamma+\symsubsup{n}{j1}{i},\dots,\gamma + \symsubsup{n}{jj-1}{i},\gamma + \kappa + \symsubsup{n}{jj}{i},\gamma + \symsubsup{n}{jj+1}{i},\dots]\otimes \fset{i}).\label{eqn:piPosterior} 
\end{align}
Here, $\symsubsup{n}{jk}{i}$ are the number of transitions from mode $j$ to $k$ in $\symsubsup{z}{1:T}{i}$. Since the mode sequence $\symsubsup{z}{1:T}{i}$ was generated from feature-constrained transition distributions, $\symsubsup{n}{jk}{i}$ will be zero for any $k$ such that $f_{ik}=0$. Using the definition of $\symsubsup{\pi}{j}{i}$ in Eq.~\eqref{eqn:normEta}, one can equivalently define a sample from the posterior of Eq.~\eqref{eqn:piPosterior} by solely updating $\symsubsup{\eta}{jk}{i}$ for instantiated features:
\begin{align}
	\symsubsup{\eta}{jk}{i}\mid \symsubsup{z}{1:T}{i},\gamma,\kappa \sim \mbox{Gamma}(\gamma+\kappa\delta(j,k)+\symsubsup{n}{jk}{i},1), \quad k \in \{\ell \mid f_{i\ell}=1\}. 
\end{align}
\subsubsection*{Dynamic parameters $\{\BF{A}_k,\Sigma_k\}$} We now turn to posterior updates for dynamic parameters.  Recall the conjugate matrix normal inverse-Wishart (MNIW) prior on $\{\BF{A}_k,\Sigma_k\}$, comprised of an inverse-Wishart prior $\mbox{IW}(n_0,S_0)$ on $\Sigma_k$ and a matrix-normal prior $\MN{\BF{A}_k}{M}{\Sigma_k}{K}$ on $\BF{A}_k$ given $\Sigma_k$.  We consider the following sufficient statistics based on the sets $\BF{Y}_{\!k} = \{\symsubsupB{y}{t}{i} \mid \symsubsup{z}{t}{i} = k, \, i=1,\ldots,N\}$ and \mbox{$\BF{\tilde{Y}}_{\!k} = \{\symsubsupB{\tilde{y}}{t}{i} \mid \symsubsup{z}{t}{i} = k, \, i=1,\ldots,N\}$} of observations and lagged observations, respectively, associated with behavior $k$:
\begin{equation}
	\begin{aligned}
		\symsubsup{S}{\tilde{y}\tilde{y}}{k} = \sum_{(t,i)\mid \symsubsup{z}{t}{i} = k} \symsubsupB{\tilde{y}}{t}{i}\symsubsupT{\tilde{y}}{t}{i} + \BF{K} &\hspace{0.25in} \symsubsup{S}{y\tilde{y}}{k} = \sum_{(t,i)\mid \symsubsup{z}{t}{i} = k} \symsubsupB{y}{t}{i}\symsubsupT{\tilde{y}}{t}{i} + \BF{M}\BF{K}\\
		\symsubsup{S}{yy}{k} = \sum_{(t,i)\mid \symsubsup{z}{t}{i} = k} \symsubsupB{y}{t}{i}\symsubsupT{y}{t}{i} + \BF{M}\BF{K}\BF{M}^T &\hspace{0.25in} \symsubsup{S}{y|\tilde{y}}{k} = \symsubsup{S}{yy}{k} - \symsubsup{S}{y\tilde{y}}{k}S_{\tilde{y}\tilde{y}}^{-(k)}S_{\tilde{y}\tilde{y}}^{(k)^T}. 
	\end{aligned}
	\label{eqn:Sk} 
\end{equation}
It is through this pooling of data from multiple time series that we improve our inferences on shared behaviors, especially in the presence of limited data. Using standard MNIW conjugacy results, the posterior can be shown to equal
\begin{equation}
	\begin{aligned}
		\BF{A}_k \mid \Sigma_k,\BF{Y}_{\!k} &\sim \MN{\BF{A}_k}{\symsubsup{S}{y\tilde{y}}{k}S_{\tilde{y}\tilde{y}}^{-(k)}}{\Sigma_k}{\symsubsup{S}{\tilde{y}\tilde{y}}{k}}\\
		\Sigma_k \mid \BF{Y}_{\!k} &\sim \mbox{IW}\left(|\BF{Y}_{\!k}| + n_0, \symsubsup{S}{y|\tilde{y}}{k} + S_0\right). 
	\end{aligned}
\end{equation}
\subsection{Sampling the BP and Dirichlet transition hyperparameters} \label{sec:IBPhyperparameters}
We additionally place priors on the Dirichlet hyperparameters $\gamma$ and $\kappa$, as well as the BP parameter $\alpha$.

\subsubsection*{BP hyperparameter $\alpha$} Let $\BF{F}=\{\BF{f}_i\}$. As derived by~\citet{GriffithsGhahramani:05}, $p(\BF{F} \mid \alpha)$ can be expressed as
\begin{align}
	p(\BF{F} \mid \alpha) \propto \alpha^{K_+}\exp\bigg(-\alpha\sum_{n=1}^N \frac{1}{n}\bigg), 
\end{align}
where, as before, $K_+$ is the number of unique features activated in $\BF{F}$. As in~\citet{Gorur:06}, we place a conjugate $\mbox{Gamma}(a_\alpha,b_\alpha)$ prior on $\alpha$, which leads to the following posterior distribution:
\begin{align}
	p(\alpha \mid \BF{F},a_\alpha,b_\alpha) &\propto \alpha^{K_+}\exp\left(-\alpha\sum_{n=1}^N \frac{1}{n}\right) \cdot \frac{\alpha^{a_\alpha-1}\exp(-b_\alpha \alpha)}{\Gamma(\alpha)}\nonumber\\
	&=\mbox{Gamma}\bigg(a_\alpha+K_+,b_\alpha+\sum_{n=1}^N \frac{1}{n}\bigg) 
\end{align}
\subsubsection*{Transition hyperparameters $\gamma$ and $\kappa$} Transition hyperparameters are assigned priors $\gamma \sim \mbox{Gamma}(a_\gamma,b_\gamma)$ and $\kappa \sim \mbox{Gamma}(a_\kappa,b_\kappa)$. Because the generative process of Eq.~\eqref{eqn:transitionGamma} is non-conjugate, we rely on Metropolis-Hastings steps which iteratively sample $\gamma$ given $\kappa$, and $\kappa$ given $\gamma$. Each sub-step uses a gamma proposal distribution $q_\gamma(\cdot\mid\cdot)$ or $q_\kappa(\cdot\mid\cdot)$, respectively, with fixed variance $\sigma_\gamma^2$ or $\sigma_\kappa^2$, and mean equal to the current hyperparameter value. 

As derived in Appendix~\ref{app:transparams}, the acceptance ratio for for the proposal of $\gamma$ given $\kappa$ is
\begin{align}
	r(\gamma' \mid \gamma) = \frac{f(\gamma')\Gamma(\vartheta)\gamma^{\vartheta'-\vartheta-a_\gamma}}{f(\gamma)\Gamma(\vartheta')\gamma'^{\vartheta-\vartheta'-a_\gamma}} \exp\{-(\gamma' - \gamma)b_\gamma\} \sigma_\gamma^{2(\vartheta-\vartheta')},
\end{align}
where $\vartheta=\gamma^2/\sigma_{\gamma}^2$, $\vartheta'=\gamma'^2/\sigma_{\gamma}^2$, and $f(\gamma)$ is the likelihood term.  Specifically, letting $\BF{\pi} = \{\pi_j^{(i)}\}$ and recalling the definition of $\symsubsup{\tilde{\pi}}{j}{i}$ from Eq.~\eqref{eqn:DirPrior} and that $K_i = \sum_k f_{ik}$, the likelihood term may be written as
\begin{align}
	f(\gamma) \triangleq p(\BF{\pi}\mid \gamma,\kappa,\BF{F}) = \prod_i \prod_{k=1}^{K_i} \left\{\frac{\Gamma(\gamma K_i + \kappa)}{\left(\prod_{j=1}^{K_i-1} \Gamma(\gamma)\right)\Gamma(\gamma+\kappa)} \prod_{j=1}^{K_i} \tilde{\pi}_{kj}^{(i)^{\gamma+\kappa\delta(k,j)-1}}\right\}. 
\end{align}

The Metropolis-Hastings sub-step for sampling $\kappa$ given $\gamma$ follows similarly. In this case, however, the likelihood terms simplifies to
\begin{align}
	f(\kappa) \triangleq \prod_i \frac{\Gamma(\gamma K_i + \kappa)^{K_i}}{\Gamma(\gamma+\kappa)^{K_i}} \prod_{j=1}^{K_i} \tilde{\pi}_{jj}^{(i)^{\gamma+\kappa-1}} \propto p(\BF{\pi}\mid \gamma,\kappa,\BF{F}). 
\end{align}
The resulting MCMC sampler for the BP-AR-HMM is summarized in Algorithm~\ref{alg:IBPARHMMsampler} of Appendix~\ref{app:alg}.
\section{Related Work}
\label{sec:related}
A challenging problem in deploying Markov switching processes such as the AR-HMM is that of defining the number of dynamic regimes.  Previously, Bayesian nonparametric approaches building on the hierarchical Dirichlet process (HDP)~\citep{Teh:06} have been proposed to allow uncertainty in the number of regimes by defining Markov switching processes on infinite state spaces~\citep{Beal:02,Teh:06,Fox:AOAS11,Fox:IEEE11}.  See~\citet{Fox:IEEESPM} for a recent review.  However, these formulations focus on a single time series whereas in this paper our motivation is analyzing a collection of time series.  A na\"{i}ve approach to employing such models in the multiple time series setting is to simply couple each of the time series under a shared HDP prior.  However, such an approach assumes that the state spaces of the multiple Markov switching processes are \emph{exactly} shared, as are the transitions among these states (i.e., both the transition and emissions parameters are global.)  As demonstrated in Section~\ref{sec:synth} and Section~\ref{sec:MoCap}, such strict sharing can limit the ability to discover unique dynamic behaviors and reduce the predictive performance of the inferred model. 

In recent independent work, \citet{saria2010discovering} developed an alternative approach to modeling multiple time series via the HDP-HMM.  Their \emph{time series topic model} (TSTM) describes coarse-scale temporal behavior using a finite set of ``topics'', which are themselves distributions on a common set of autoregressive dynamical models.  Each time series is assumed to exhibit all topics to some extent, but with unique frequencies and temporal patterns.
Alternatively, the mixed HMM~\citep{altman2007mixed} uses generalized linear models to allow the state transition and emission distributions of a finite HMM to depend on arbitrary external covariates.  In experiments, this is used to model the differing temporal dynamics of a small set of known time series classes.

More broadly, the specific problem we address here has received little previous attention, perhaps due to the difficulty of treating such combinatorial relationships with parametric models.  There are a wide variety of models which capture correlations among multiple aligned, interacting univariate time series, for example using Gaussian state space models~\citep{aoki1991state}.
Other approaches cluster time series using a parametric mixture model~\citep{alon2003discovering}, or a Dirichlet process mixture~\citep{qi07}, and model the dynamics within each cluster via independent finite HMMs.

Dynamic Bayesian networks~\citep{murphy2002dynamic}, such as the factorial HMM~\citep{ghahramani97}, define a structured representation for the latent states underlying a single time series.
Such models are widely used in applied time series analysis~\citep{lehrach2009segmenting,duh2005jointly}.
The infinite factorial HMM~\citep{VanGael:08_2} uses the IBP to model a single time series via an infinite set of latent features, each evolving according to independent Markovian dynamics. Our work instead focuses on modeling multiple time series and on capturing dynamical modes that are shared among the series.

Other approaches do not explicitly model latent temporal dynamics, and instead aim to align time series with consistent global structure~\citep{aach2001aligning}.
Motivated by the problem of detecting temporal anomalies, \citet{listgarten2007bayesian} describe a hierarchical Bayesian approach to modeling shared structure among a known set of time series classes.  Independent HMMs are used to encode non-linear alignments of observed signal traces to latent reference time series, but their states do not represent dynamic behaviors and are not shared among time series.

\section{Synthetic Experiments}
\label{sec:synth}
\subsection{Discovering Common Dynamics}
\begin{figure}
	[t!] \centering 
	\begin{tabular}
		{c} \hspace{-0.1in}
		\includegraphics[height = 2in]{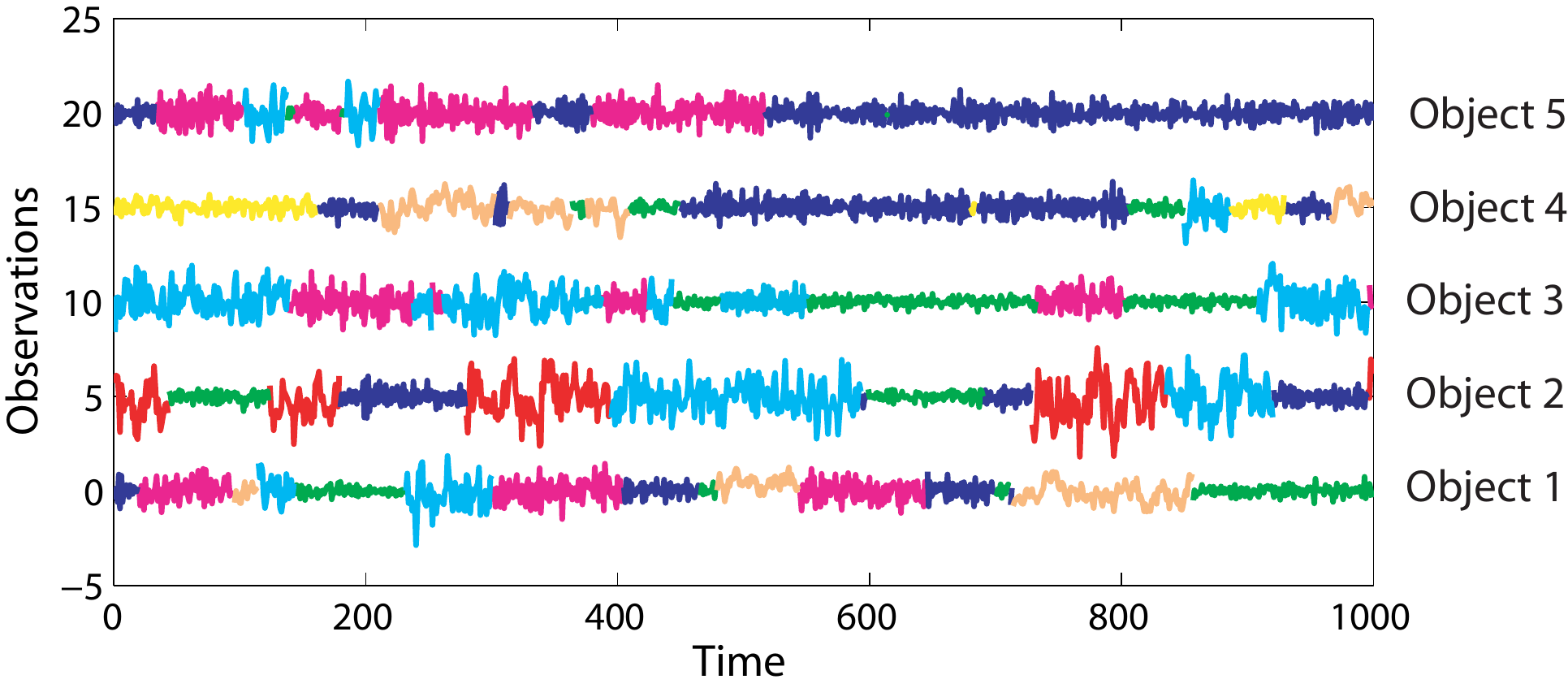}\\
		(a) 
	\end{tabular}
	\vspace{0.05in} 
	\begin{tabular}
		{cc} 
		\includegraphics[height = 1.5in]{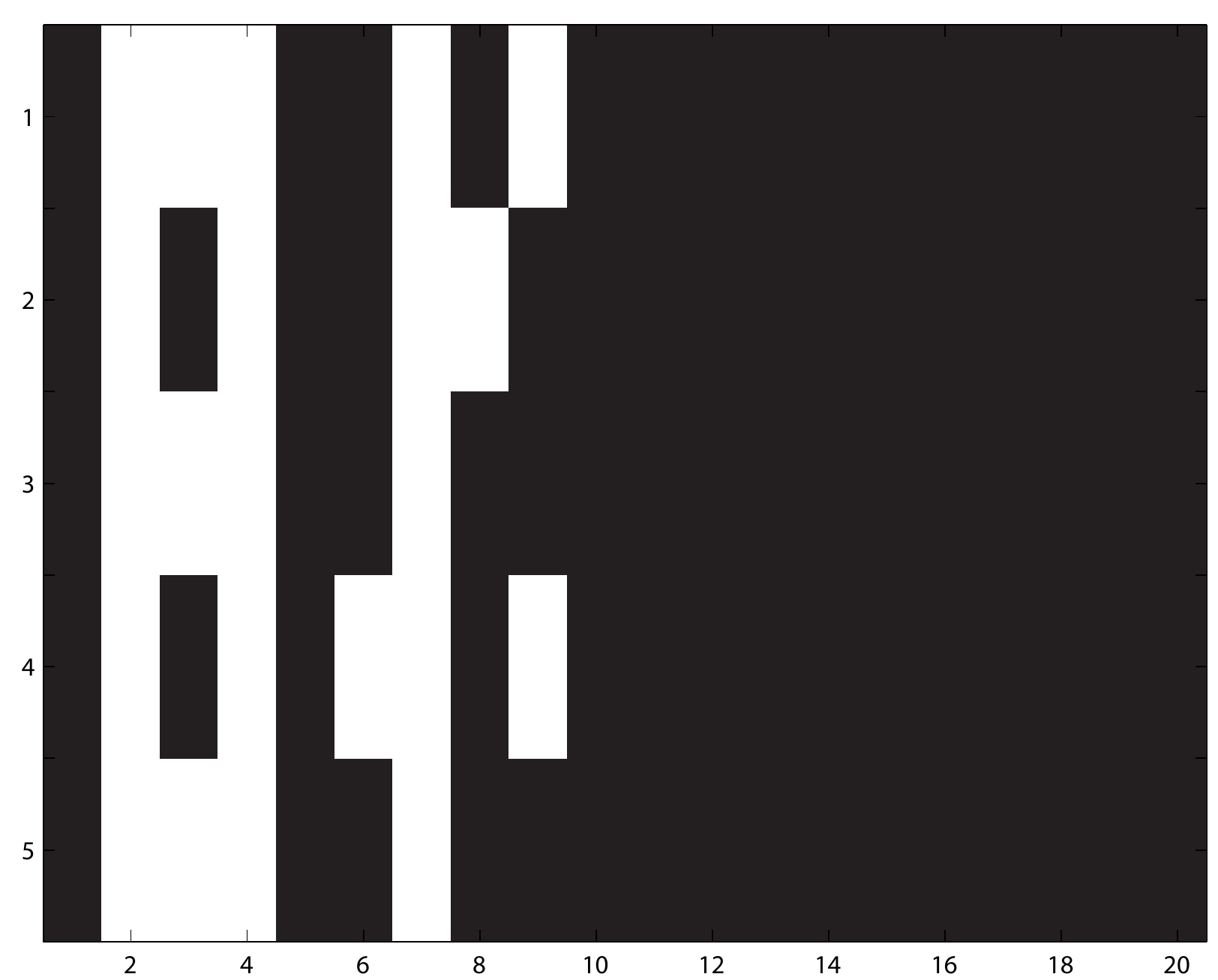} & 
		\includegraphics[height = 1.5in]{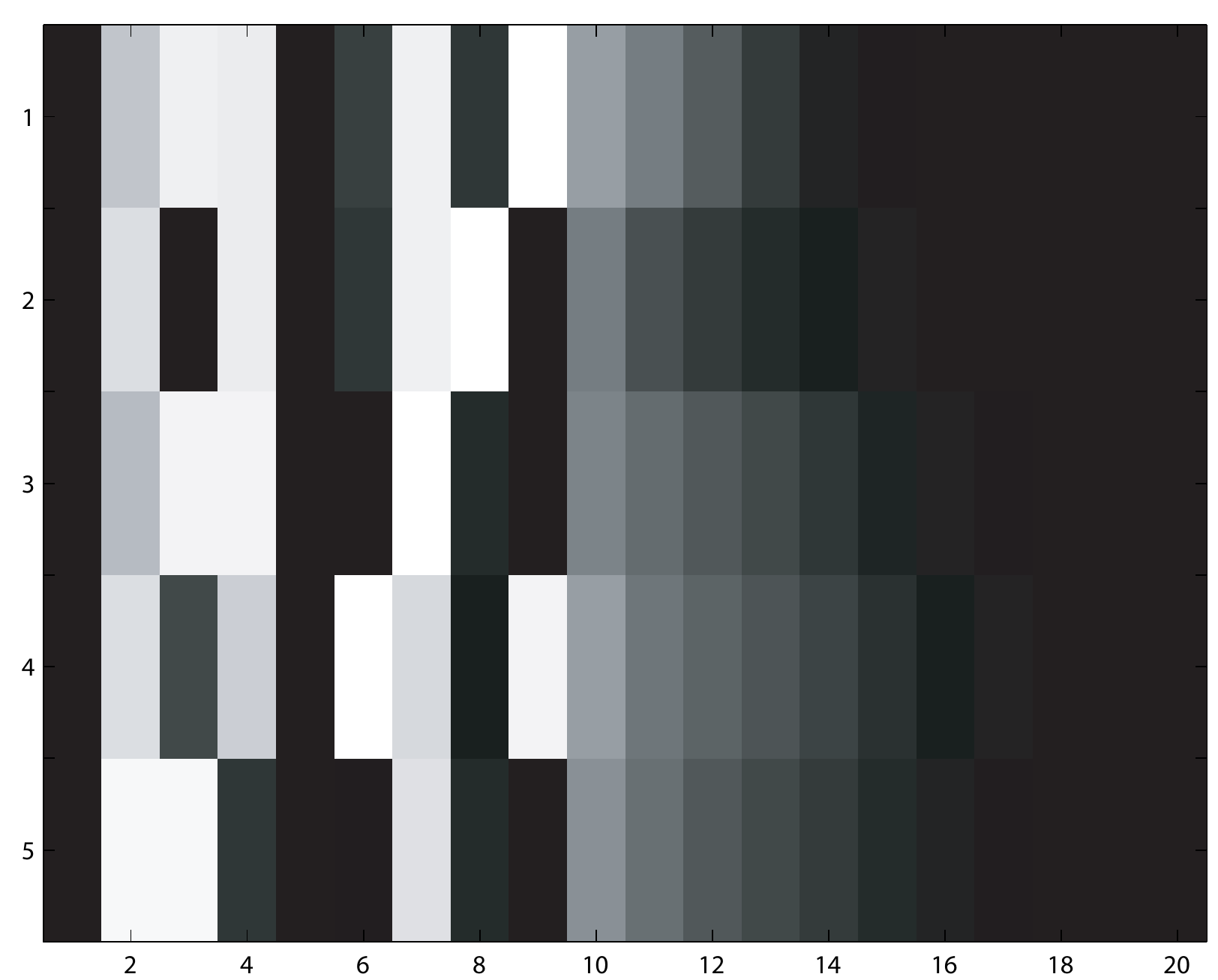}\\
		(b) & (c)		
	\end{tabular}
	\caption[Synthetic data for 5 switching AR(1) time series, and associated true and BP-AR-HMM learned feature matrices.]{(a) Observation sequences for each of 5 switching AR(1) time series colored by true mode sequence, and offset for clarity. Images of the \mbox{(b)} true feature matrix of the five time series and (c) estimated feature matrix averaged over 10,000 MCMC samples taken from 100 trials every 10th sample. Each row corresponds to a different time series, and each column a different autoregressive model. White indicates active features. Although the true model is defined by only 9 possible dynamical modes, we show 20 columns in order to display the ``tail'' of the BP-AR-HMM estimated matrix resulting from samples that incorporated additional dynamical modes (events that have positive probability of occurring, as defined by the IBP prior.) The estimated feature matrices are produced from mode sequences mapped to the ground truth labels according to the minimum Hamming distance metric, and selecting modes with more than 2\% of the observations in a time series.} \label{fig:results1} 
\end{figure}

To test the ability of the BP-AR-HMM to discover shared dynamics, we generated five time series that switched between AR(1) models:
\begin{align}
	\symsubsup{y}{t}{i} = a_{\symsubsup{z}{t}{i}}\symsubsup{y}{t-1}{i}+\symsubsup{e}{t}{i}(\symsubsup{z}{t}{i}),
\end{align}
with $a_k \in \{-0.8,-0.6,-0.4,-0.2,0,0.2,0.4,0.6,0.8\}$ and process noise covariance $\Sigma_k$ drawn from an $\mbox{IW}(3,0.5)$ prior. The time-series-specific features, shown in Fig.~\ref{fig:results1}(b), were sampled from a truncated IBP~\citep{GriffithsGhahramani:05} using $\alpha=10$ 
and then used to generate the observation sequences of Fig.~\ref{fig:results1}(a) (colored by the true mode sequences). Each row of the feature matrix corresponds to one of the five time series, and the columns represent the different autoregressive models with a white square indicating that a given time series uses that dynamical mode. Here, the columns are ordered so that the first feature corresponds to an autoregressive model defined by $a_1$, and the ninth feature corresponds to that of $a_9$.
%

The resulting feature matrix estimated over 10,000 MCMC samples is shown in Fig.~\ref{fig:results1}(c). Each of the 10,000 estimated feature matrices is produced from an MCMC sample of the mode sequences that are first mapped to the ground truth labels according to the minimum Hamming distance metric. We then only maintain inferred dynamical modes with more than 2\% of the time series's observations. Comparing to the true feature matrix, we see that our model is indeed able to discover most of the underlying latent structure of the time series despite the challenges caused by the fact that the autoregressive coefficients are close in value. The most commonly missed feature occurrence is the use of $a_4$ by the fifth time series. This fifth time series is the top-most displayed in Fig.~\ref{fig:results1}(a), and the dynamical mode defined by $a_4$ is shown in green. We see that this mode is used very infrequently, making it challenging to distinguish. Due to the nonparametric nature of the model, we also see a ``tail'' in the estimated matrix because of the (infrequent) incorporation of additional dynamical modes.

\subsection{Comparing the Feature-Based Model to Nonparametric Models with Identical State Spaces}
One might propose, as an alternative to the BP-AR-HMM, the use of an architecture based on the hierarchical Dirichlet process of~\citet{Teh:06}; specifically we could use the HDP-AR-HMMs of~\citet{Fox:IEEE11} tied together with a shared set of transition and dynamic parameters. For an HDP-AR-HMM truncated to $L$ possible dynamical modes, this model is specified as:
\begin{equation}
	\begin{aligned}
		\beta &\sim \mbox{Dir}(\gamma/L,\ldots,\gamma/L)\\
		\pi_j \mid \beta &\sim \mbox{Dir}(\alpha\beta_1,\dots,\alpha\beta_{j-1},\alpha\beta_j + \kappa,\alpha\beta_{j+1},\dots,\alpha\beta_L)\\
		\symsubsup{z}{t}{i} &\sim \pi_{\symsubsup{z}{t-1}{i}}, \quad 
		\symsubsupB{y}{t}{i} = \BF{A}_{\symsubsup{z}{t}{i}}\symsubsupB{\tilde{y}}{t}{i} + \symsubsupB{e}{t}{i}(\symsubsup{z}{t}{i}).
	\end{aligned}
\end{equation}
Here, $\alpha$ and $\gamma$ are a set of concentration parameters that define the HDP and $\kappa$ is the sticky hyperparameter of the sticky HDP-HMM~\citep{Fox:AOAS11}; these hyperparameters are often given priors as well.  

\subsubsection*{Segmentation Performance}
\begin{figure}[p!] 
\centering 
	\begin{tabular}
		{c} 
		\includegraphics[width = 0.85\columnwidth]{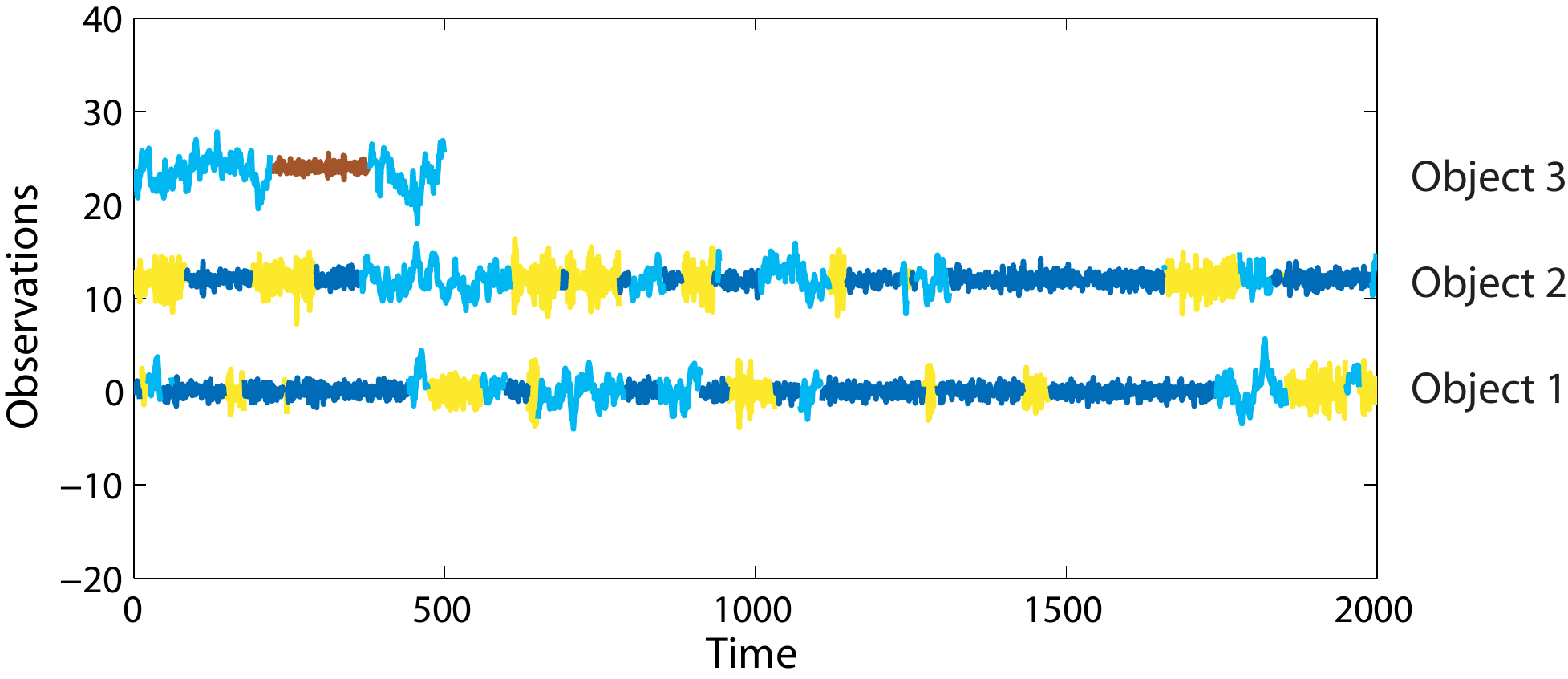}\\
		(a) 
	\end{tabular}
	\begin{tabular}
		{cc} \hspace{-0.1in}
		\includegraphics[height = 1.75in]{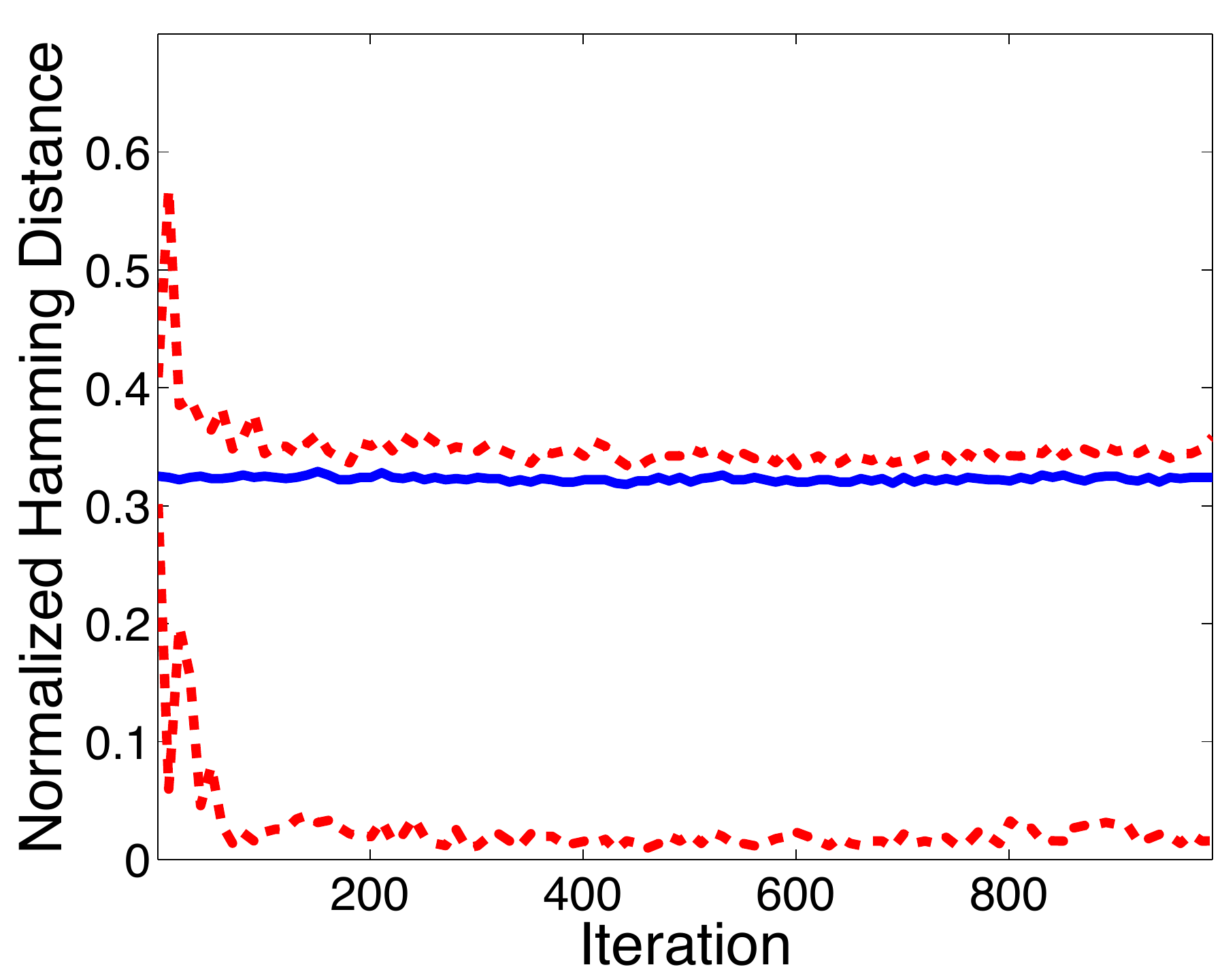} & 
		\includegraphics[height = 1.75in]{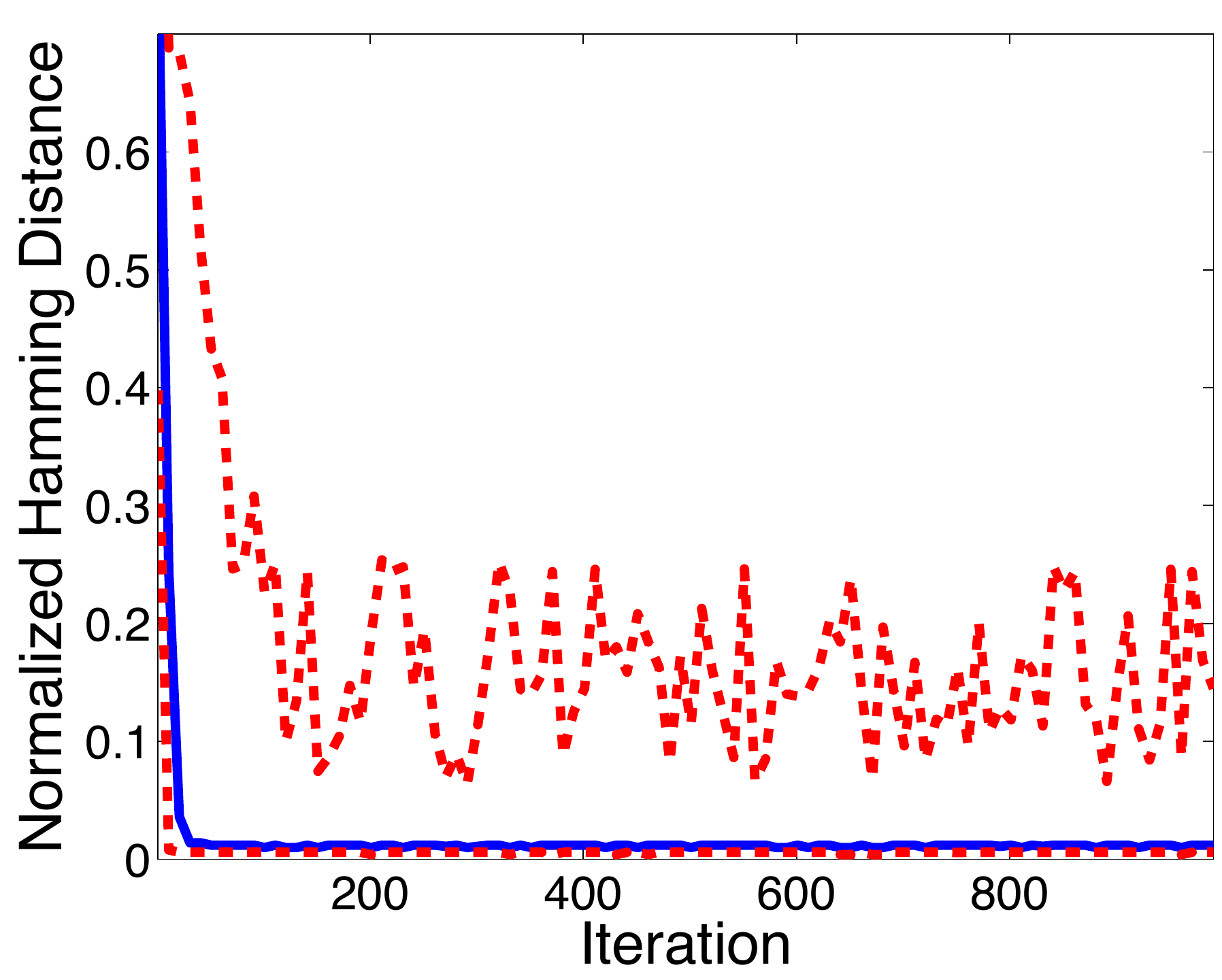}\vspace{-0.05in}\\
		(b) & (c)\vspace{0.1in}\\
		\includegraphics[width = 2.2in]{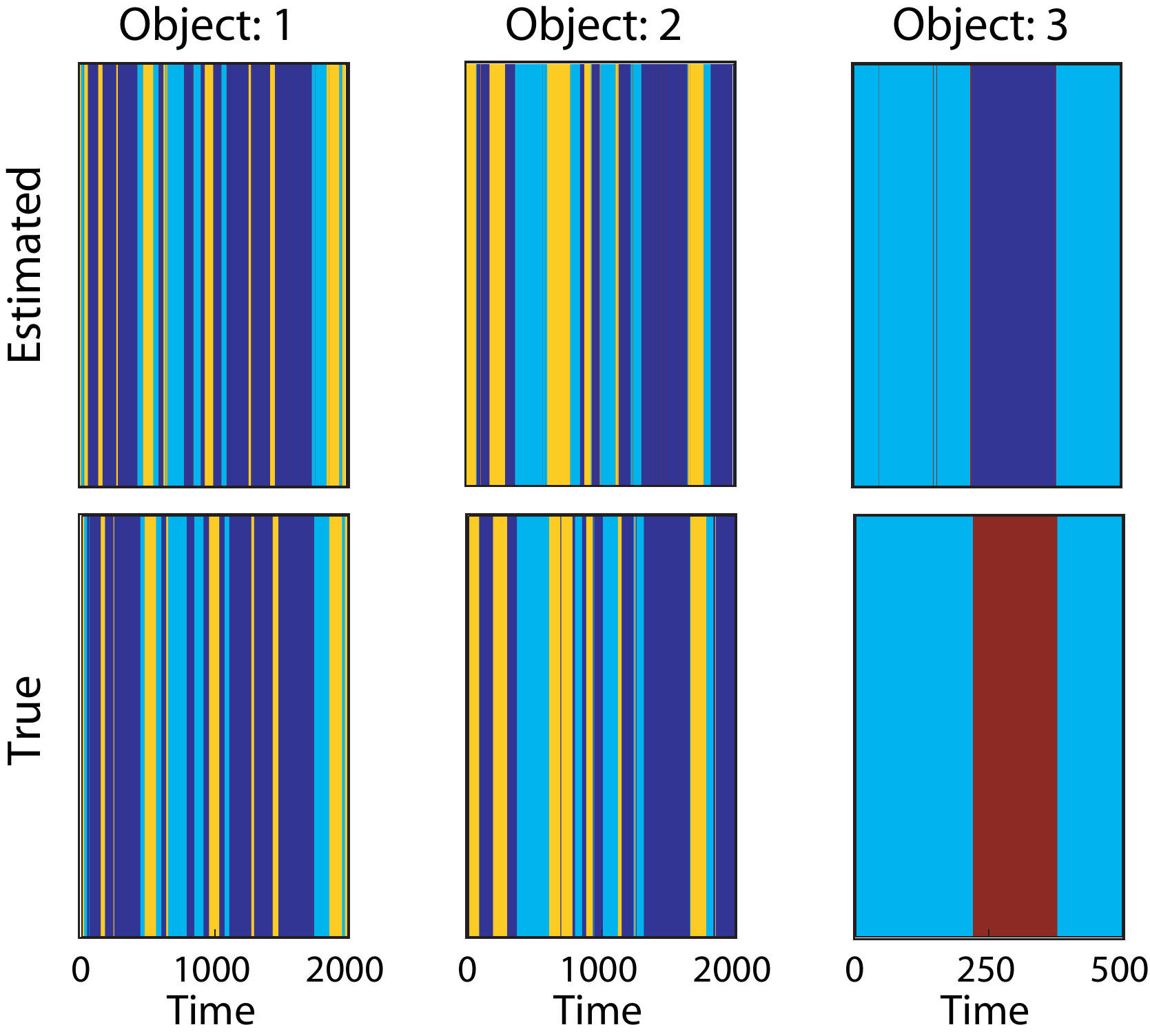}\hspace{0.05in} &\hspace{0.05in} 
		\includegraphics[width = 2.2in]{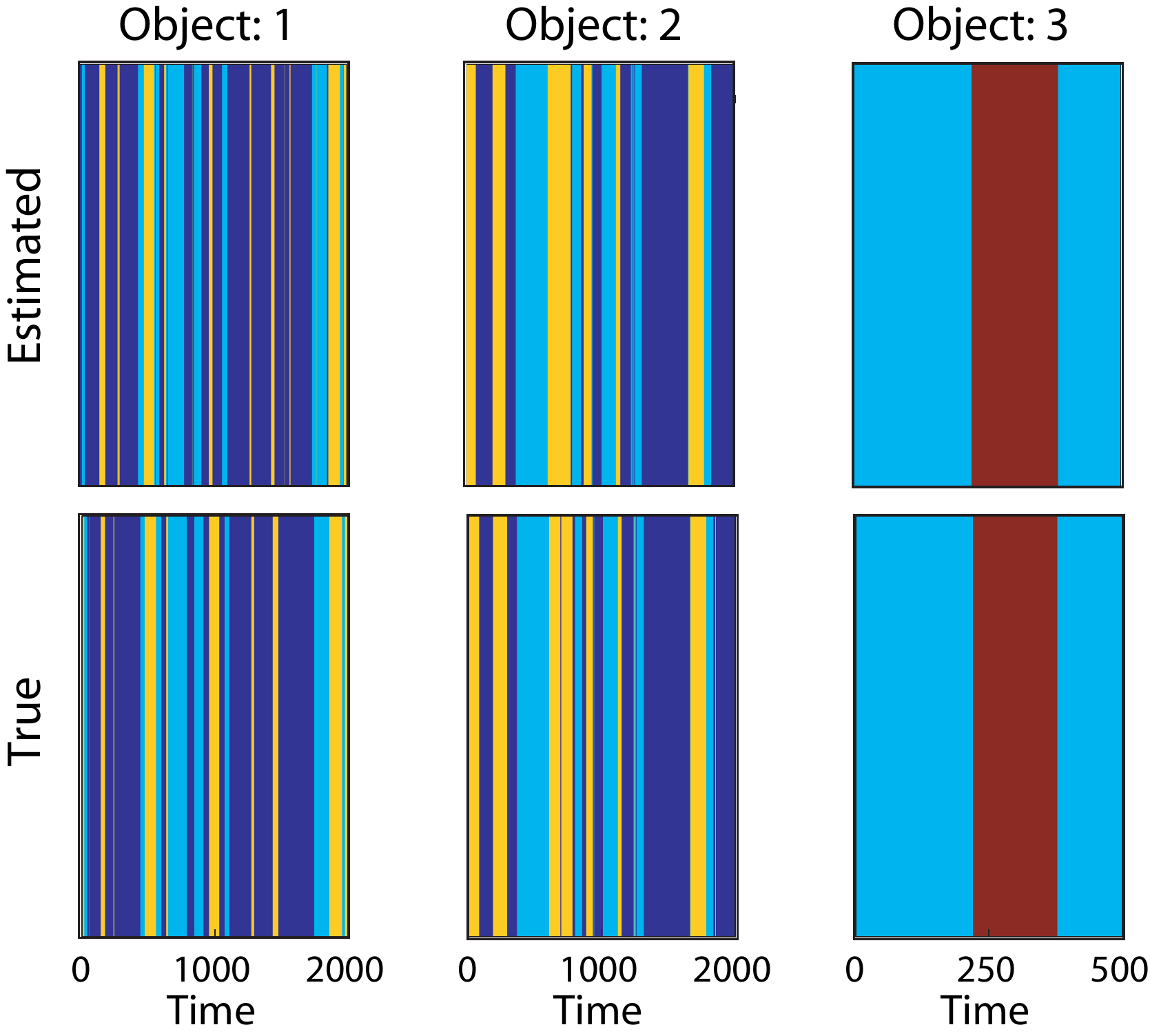}\vspace{-0.05in}\\
		(d) & (e) 
	\end{tabular}
	\caption[Hamming distance quantiles comparing the segmentation performance of the HDP-AR-HMM to the BP-AR-HMM on a synthetic data example.]{(a) Observation sequences for each of 3 switching AR(1) time series colored by true mode sequence, and offset for clarity. The first and second sequences are four times as long as the third. (b)-(c) Focusing solely on the third time series, the median (solid blue) and $10^{th}$ and $90^{th}$ quantiles (dashed red) of Hamming distance between the true and estimated mode sequence over 1000 trials are displayed for the multiple HDP-AR-HMM model~\citep{Fox:IEEE11} and the BP-AR-HMM, respectively. (d)-(e) Examples of typical segmentations into behavior modes for the three time series at MCMC iteration 1000 for the two models. The top and bottom panels display the estimated and true sequences, respectively, and the color coding corresponds exactly to that of (a). For example, time series 3 switches between two modes colored by cyan and maroon.} \label{fig:results2} 
\end{figure}

To demonstrate the difference between this HDP-AR-HMM and the BP-AR-HMM, we generated data for three switching AR(1) processes. The first two time series, with four times the data points of the third, switched between dynamical modes defined by $a_k \in \{-0.8, -0.4, 0.8\}$ and the third time series used $a_k \in \{-0.3, 0.8\}$. The results shown in Fig.~\ref{fig:results2} indicate that the multiple HDP-AR-HMM model, which assumes all time series share \emph{exactly} the same transition matrices and dynamic parameters, typically describes the third time series using $a_k \in \{-0.4, 0.8\}$ since this assignment better matches the parameters defined by the other (lengthy) time series. This common grouping of two distinct dynamical modes leads to the large median and 90th Hamming distance quantiles shown in Fig.~\ref{fig:results2}(b). The BP-AR-HMM, on the other hand, is better able to distinguish these dynamical modes (see Fig.~\ref{fig:results2}(c)) since the penalty in not sharing a behavior is only in the feature matrix; once a unique feature is chosen, it does not matter how the time series chooses to use it. Example segmentations representative of the median Hamming distance error are shown in Fig.~\ref{fig:results2}(d)-(e). These results illustrate that the IBP-based feature model emphasizes choosing behaviors rather than assuming all time series are performing minor variations of the same dynamics.

For the experiments above, we placed a $\mbox{Gamma}(1,1)$ prior on $\alpha$ and $\gamma$, and a $\mbox{Gamma}(100,1)$ prior on $\kappa$. The gamma proposals used $\sigma_{\gamma}^2=1$ and $\sigma_{\kappa}^2=100$ while the MNIW prior was given $M=0$, $K=0.1*I_d$, $n_0 = d+2$, and $S_0$ set to 0.75 times the empirical variance of the joint set of first-difference observations. At initialization, each time series was segmented into five contiguous blocks, with feature labels unique to that sequence.

\subsubsection*{Predictive Performance}
\begin{figure}[t!] 
	\centering  
		\includegraphics[width = 2.75in]{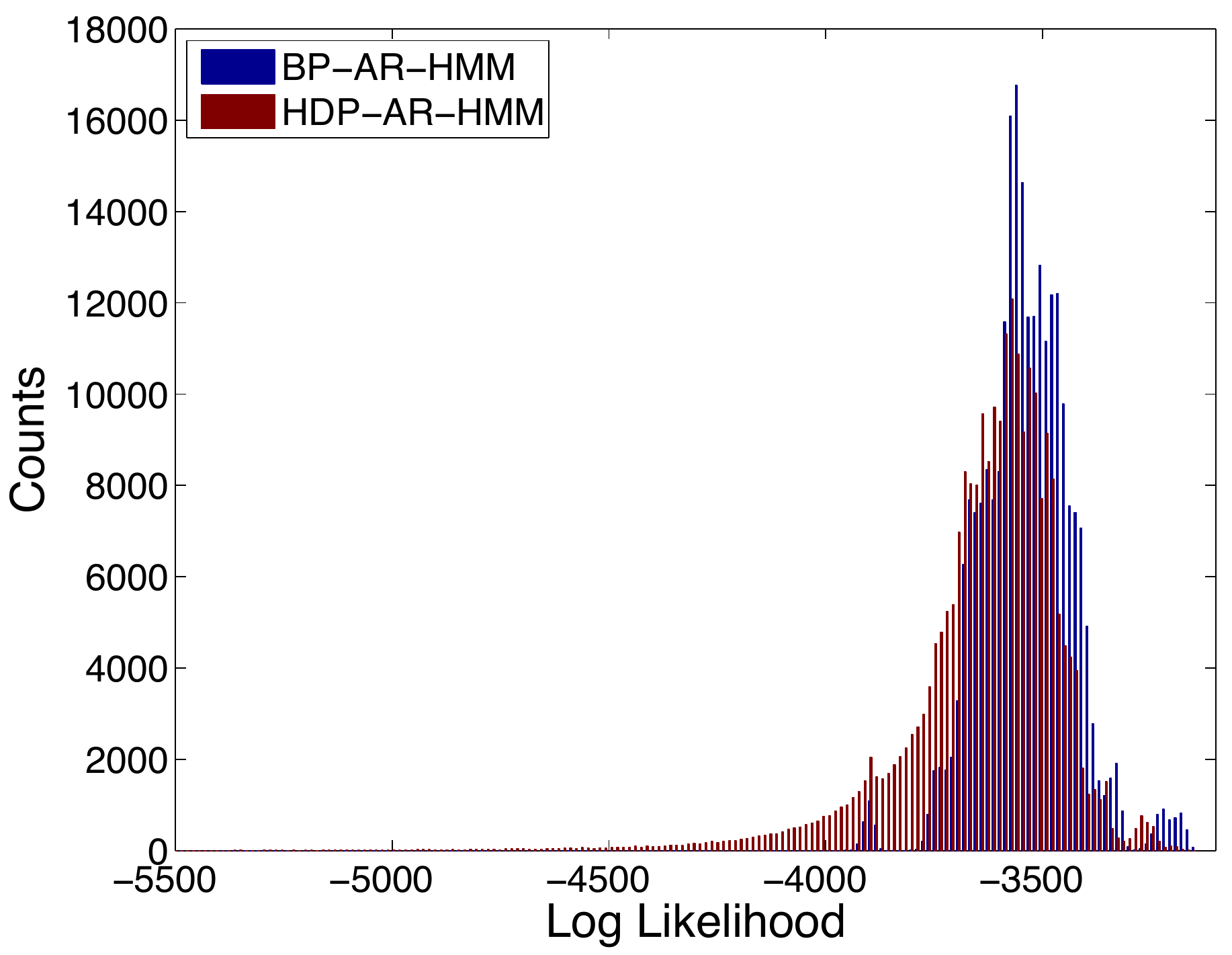} 
		\caption{Histogram of the predictive log-likelihood of 100 held-out data using the inferred parameters sampled every 10th iteration from MCMC iterations 500–1,000 from 50 independent chains for the BP-AR-HMM and HDP-AR-HMM run on the data of Fig.~\ref{fig:results2}(a).} \label{fig:hist_pred_perf} 
\end{figure}

Using the same data-generating mechanism as used to generate the time series displayed in Fig.~\ref{fig:results2}(a), we generated a set of 100 held-out test datasets for Objects 1, 2, and 3.  Each of the time series comprising the test datasets was of length 1000 (in contrast to the data of Fig.~\ref{fig:results2}(a) in which the time series of Object 3 was of length 500 and those of Objects 1 and 2 were of length 2000.)  Based on a set of samples taken from 50 chains at MCMC iterations $[500~:~10:~1000]$ (i.e., a total of 2500 samples), we computed the log-likelihood of each of the 100 held-out datasets.  That is, we added the time-series-specific log-likelihoods computed for each time series since the time series are conditionally independent given the model parameters.  We performed this task for both the MCMC samples of the BP-AR-HMM and HDP-AR-HMM.  The results are summarized in the histogram of Fig.~\ref{fig:hist_pred_perf}.

Since the BP-AR-HMM consistently identifies the unique dynamical mode of $a_k = -0.3$ used by Object 3 while the HDP-AR-HMM does not, we see from Fig.~\ref{fig:hist_pred_perf} that the mass of the BP-AR-HMM predictive log-likelihood is shifted positively by roughly 100 compared to that of the HDP-AR-HMM.  In addition, we see that the histogram for the HDP-AR-HMM has a heavy tail, skewed towards lower log-likelihood, whereas the BP-AR-HMM does not.

Recall a couple of key differences between the BP-AR-HMM and HDP-AR-HMM.  Both the HDP-AR-HMM and BP-AR-HMM define global libraries of infinitely many possible dynamic behaviors.  However, the HDP-AR-HMM assumes that each of the time series selects the same finite subset of behaviors and transitions between them in exactly the same manner (i.e., the transition matrix is also global.)  On the other hand, the BP-AR-HMM allows each time series to select differing subsets of behaviors \emph{and} differing transition probabilities.  In the dataset examined here, the data-generating transition matrix between behaviors is the same for all time series, which matches the assumption of the HDP-AR-HMM.  Second, two of the three time series share exactly the same dynamical modes, which is also close to the assumed HDP-AR-HMM formulation.  The only aspect of the data that is better modeled apriori by the BP-AR-HMM is the unique dynamical mode of Object 3.  However, there is not a large difference between this unique dynamic of $a_k = -0.3$ and the HDP-AR-HMM assumed $a_k = -0.4$.  Regardless of the fact that the data are a close fit to the assumptions made by the HDP-AR-HMM, the improved predictive log-likelihood of the BP-AR-HMM illustrates the benefits of this more flexible framework.
\section{Motion Capture Experiments} 
\label{sec:MoCap}
The linear dynamical system is a common model for describing simple human motion~\citep{Hsu:05}, and the switching linear dynamical system (SLDS) has been successfully applied to the problem of human motion synthesis, classification, and visual tracking~\citep{Pavlovic:99,Pavlovic:01}. Other approaches develop non-linear dynamical models using Gaussian processes~\citep{Wang:08} or based on a collection of binary latent features~\citep{Taylor:07}. However, there has been little effort in jointly segmenting and identifying common dynamic behaviors amongst a set of \emph{multiple} motion capture (MoCap) recordings of people performing various tasks. The BP-AR-HMM provides a natural way to handle this problem. One benefit of the proposed model, versus the standard SLDS, is that it does not rely on manually specifying the set of possible behaviors. As an illustrative example, we examined a set of six CMU MoCap exercise routines~\citep{CMUmocap}, three from Subject 13 and three from Subject 14. Each of these routines used some combination of the following motion categories: running in place, jumping jacks, arm circles, side twists, knee raises, squats, punching, up and down, two variants of toe touches, arch over, and a reach out stretch.
\begin{figure}[t!] 
	\centering 
	\includegraphics[width = 5.5in]{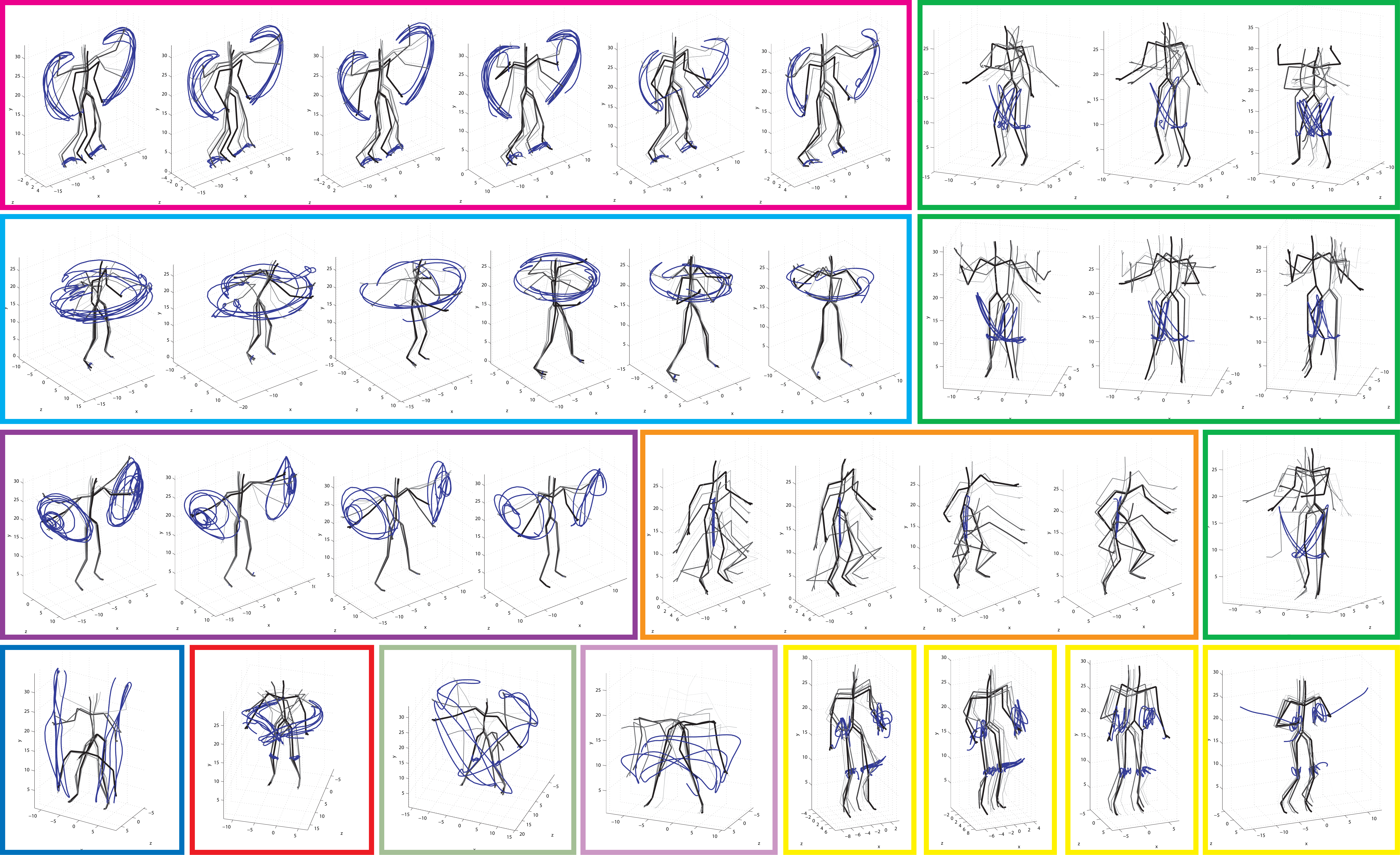} \caption[Motion capture skeleton plots for BP-AR-HMM learned segmentations of six exercise routine videos.] {Each skeleton plot displays the trajectory of a learned contiguous segment of more than two seconds. To reduce the number of plots, we preprocessed the data to bridge segments separated by fewer than 300 msec. The boxes group segments categorized under the same behavior label, with the color indicating the true behavior label (allowing for analysis of split behaviors). Skeleton rendering done by modifications to Neil Lawrence's Matlab MoCap toolbox~\citep{LawrenceMoCap}.}\label{fig:MoCap} 
\end{figure}

From the set of 62 position and joint angles, we selected the following set of 12 measurements deemed most informative for the gross motor behaviors we wish to capture: one body torso position, two waist angles, one neck angle, one set of right and left shoulder angles, the right and left elbow angles, one set of right and left hip angles, and one set of right and left ankle angles. The CMU MoCap data are recorded at a rate of at 120 frames per second, and as a preprocessing step we block-average and downsample the data using a window size of 12. We additionally scale each component of the observation vector so that the empirical variance on the concatenated set of first difference measurements is equal to one. Using these measurements, the prior distributions were set exactly as in the synthetic data experiments except the scale matrix, $S_0$, of the MNIW prior which was set to $5\cdot I_{12}$ (i.e., five times the empirical covariance of the preprocessed first-difference observations, and maintaining only the diagonal.) This setting allows more variability in the observed behaviors. We ran 25 chains of the sampler for 20,000 iterations and then examined the chain whose segmentation minimized an expected Hamming distance to the set of segmentations from all chains over iterations 15,000 to 20,000. This method of selecting a sample, first introduced in~\citet{Fox:AOAS11}, is outlined as follows.  We first choose a large reference set $\mathcal{R}$ of state sequences produced by the MCMC sampler and a possibly smaller set of test sequences $\mathcal{T}$. Then, for each collection of state sequences $\BF{z}^{[n]}$ in the test set $\mathcal{T}$ (with $\BF{z}^{[n]}$ being the MCMC sample of $\BF{z}=\{z_{1:T}^{(i)}\}$ at iteration $n$), we compute the empirical mean Hamming distance between the test sequence and the sequences in the reference set $\mathcal{R}$; we denote this empirical mean by $\hat{H}_n$. We then choose the test sequence $\BF{z}^{[n^*]}$ that minimizes this expected Hamming distance. That is,
%
\begin{align*}
\BF{z}^{[n^*]} = \arg\min_{\BF{z}^{[n]} \in \mathcal{T}} \hat{H}_n.
\end{align*}
The empirical mean Hamming distance $\hat{H}_n$ is a \emph{label-invariant loss function} since it does not rely on labels remaining consistent across samples---we simply compute
\begin{align*}
\hat{H}_n = \frac{1}{|\mathcal{R}|} \sum_{\BF{z}^{[m]} \in \mathcal{R}}
\mbox{Hamm}(\BF{z}^{[n]},\BF{z}^{[m]}),
\end{align*}
where $\mbox{Hamm}(\BF{z}^{[n]},\BF{z}^{[m]})$ is the Hamming distance between sequences $\BF{z}^{[n]}$ and $\BF{z}^{[m]}$ after finding the optimal permutation of the labels in test sequence $\BF{z}^{[n]}$ to those in reference sequence $\BF{z}^{[m]}$. At a high level, this method for choosing state sequence samples aims to produce segmentations of the data that are \emph{typical} samples from the posterior.  \citet{Jasra:05} provides an overview of some related techniques to address the label-switching issue.

The resulting MCMC sample is displayed in Fig.~\ref{fig:MoCap}. Each skeleton plot depicts the trajectory of a learned contiguous segment of more than two seconds, and boxes group segments categorized under the same behavior label by our algorithm. The color of the box indicates the true behavior label. From this plot we can infer that although some true behaviors are split into two or more categories by our algorithm, the BP-AR-HMM shows a clear ability to find common motions. Specifically, the BP-AR-HMM has successfully identified and grouped examples of jumping jacks (magenta), side twists (bright blue), arm circles (dark purple), squats (orange), and various motion behaviors that appeared in only one movie (bottom left four skeleton plots.) The split behaviors shown in green and yellow correspond to the true motion categories of knee raises and running, respectively, and the splits can be attributed to the two subjects performing the same motion in a distinct manner. For the knee raises, one subject performed the exercise while slightly twisting the upper in a counter-motion to the raised knee (top three examples) while the other subject had significant side-to-side upper body motion (middle three examples). For the running motion category, the splits also tended to correspond to varying upper body motion such as running with hands in or out of sync with knees. One example (bottom right) was the subject performing a lower-body run partially mixed with an upper-body jumping jack/arm flapping motion (an obviously confused test subject.) See Section~\ref{sec:chap5discussion} for further discussion of the BP-AR-HMM splitting phenomenon.
\begin{figure}[t!] 
	\centering 
	\includegraphics[width=.497
	\textwidth]{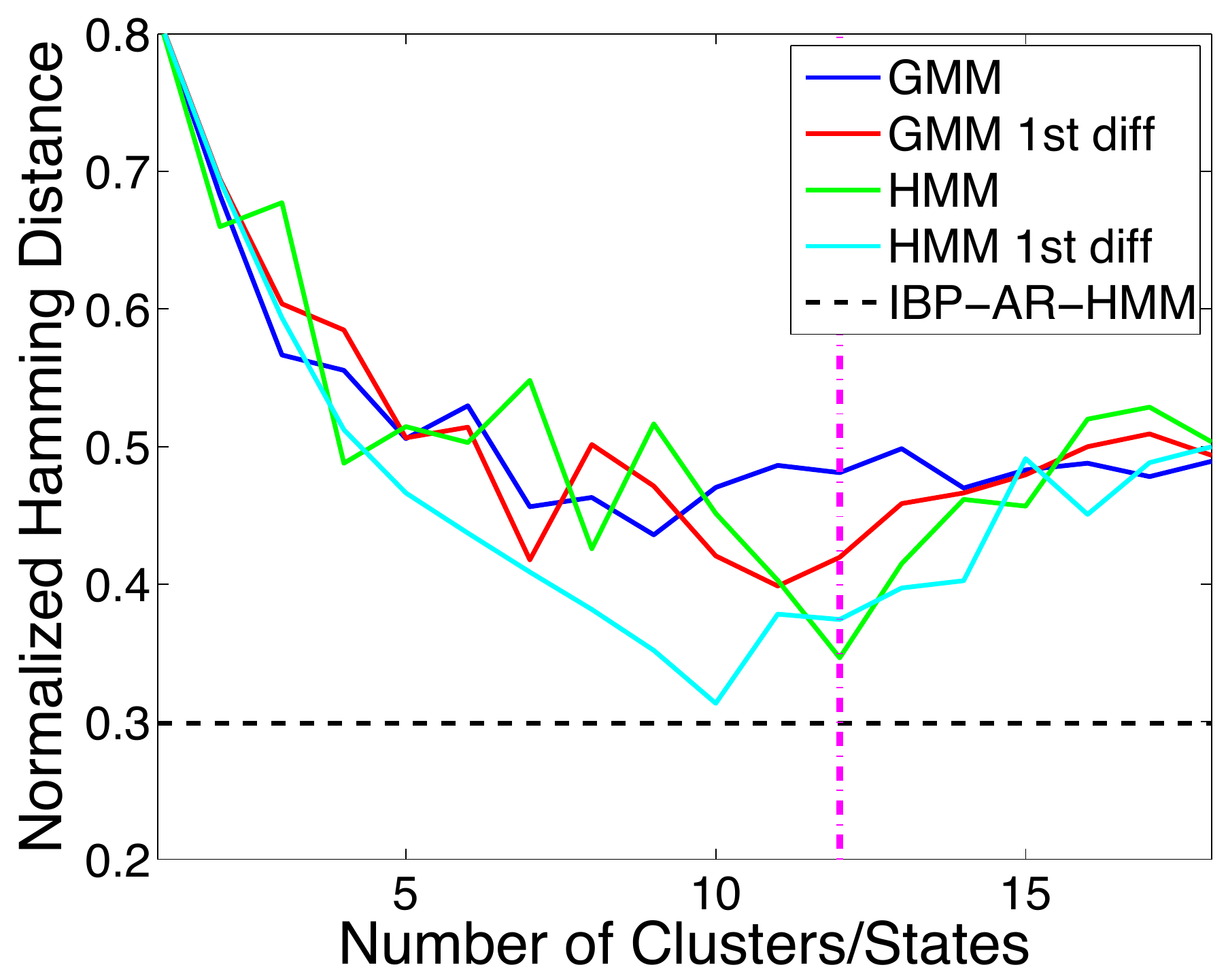} \caption[Comparison of the BP-AR-HMM MoCap segmentation performance to HMM and Gaussian mixture model approaches.]{Hamming distance versus number of GMM clusters / HMM states on raw observations (blue/green) and first-difference observations (red/cyan), with the BP-AR-HMM segmentation (black, horizontal dashed) and true feature count (magenta, vertical dashed) shown for comparison. Results are for the most-likely of ten initializations of EM using an HMM Matlab toolbox~\citep{MurphyHMMtoolbox}.} \label{fig:GMM_Hamm} 
\end{figure}
\begin{figure}
	[t!] \centering 
	\includegraphics[width=.7
	\textwidth]{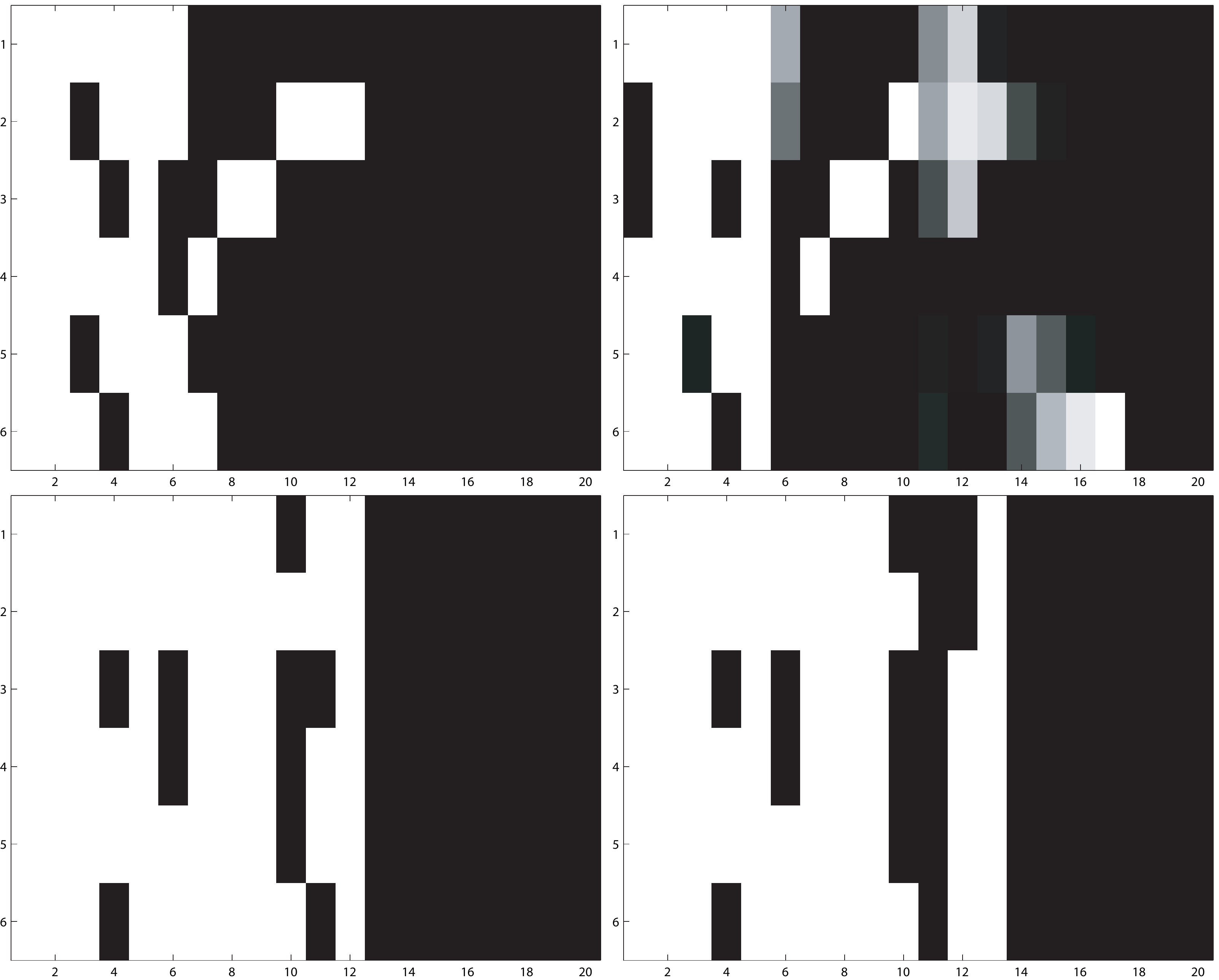} \caption[Learned MoCap feature matrices from the BP-AR-HMM, HMM, and Gaussian mixture model approaches.]{Feature matrices associated with the true MoCap sequences (top-left), BP-AR-HMM estimated sequences over iterations 15,000 to 20,000 (top-right), and MAP assignment of the GMM (bottom-left) and HMM (bottom-right) using first-difference observations and 12 clusters/states.} \label{fig:GMM_F} 
\end{figure}

We compare our MoCap performance to the Gaussian mixture model (GMM) method of~\citet{Barbic:04} using expectation maximization (EM) initialized with k-means.~\citet{Barbic:04} 
also present an approach based on probabilistic principal component analysis (PCA), but this method focuses primarily on change-point detection rather than behavior clustering. As further comparisons, we consider a GMM on first-difference observations, and an HMM on both data sets. In Fig.~\ref{fig:GMM_Hamm}, we analyze the ability of the BP-AR-HMM, as compared to the defined GMMs and HMMs, in providing accurate labelings of the individual frames of the six movie clips~\footnote{The ability to accurately label the frames of a large set of movies is useful for tasks such as querying an extensive MoCap database (such as that of CMU) without relying on manual labeling of the movies.}. Specifically, we plot the Hamming distance between the true and estimated frame labels versus the number of GMM clusters and HMM states, using the most-likely of ten initializations of EM. We also plot the Hamming distance corresponding the BP-AR-HMM MCMC sample depicted in Fig.~\ref{fig:MoCap}, demonstrating that the BP-AR-HMM provides more accurate frame labels than any of these alternative approaches over a wide range of mixture model settings. The estimated feature matrices for the BP-AR-HMM and the GMM and HMM on first difference observations are shown in Fig.~\ref{fig:GMM_F}. The figure displays the matrix associated with the MAP label estimate in the case of the GMM and HMM, and an estimate based on MCMC samples from iterations 15,000 to 20,000 for the BP-AR-HMM. For the GMM and HMM, we consider the case when the number of Gaussian mixture components or the number of HMM states is set to the true number of behaviors, namely 12. By pooling all of the data, the GMM and HMM approaches assume that each time series exhibits the same structure; the results of this assumption can be seen in the strong bands of white implying sharing of behavior between the time series. The feature matrix estimated by the BP-AR-HMM, on the other hand, provides a much better match to the true matrix by allowing for sequence-specific variability. For example, this ability is indicated by the special structure of features in the upper right portion of the true feature matrix that is mostly captured in the BP-AR-HMM estimated feature matrix, but is not present in those of the GMM or HMM. We do, however, note a few BP-AR-HMM merged and split behaviors. Overall, we see that in addition to producing more accurate segmentations of the MoCap data, the BP-AR-HMM provides a superior ability to discover the shared feature structure.
\section{Discussion} \label{sec:chap5discussion}
We have presented a Bayesian nonparametric framework for discovering dynamical modes common to multiple time series.  Our formulation reposes on the beta process, which provides a prior distribution on overlapping subsets of binary features.  This prior allows both for commonality and time-series-specific variability in the use of dynamical modes. We additionally developed a novel exact sampling algorithm for non-conjugate IBP models. The utility of our BP-AR-HMM was demonstrated both on synthetic data, and on a set of MoCap sequences where we showed performance exceeding that of alternative methods. Although we focused on switching VAR processes, our approach could be equally well applied to HMMs, and to a wide range of other segmented dynamical systems models such as switching linear dynamic systems.

The idea proposed herein of a feature-based approach to relating multiple time series is not limited to nonparametric modeling.  One could just as easily employ these ideas within a parametric model that pre-specifies the number of possible dynamic behaviors.  We emphasize, however, that conditioned on the infinite feature vectors of our BP-AR-HMM, our model reduces to a collection of Markov switching processes on a \emph{finite} state space.  The beta process simply allows for flexibility in the overall number of globally shared behaviors, and computationally we do not rely on any truncations of this infinite model.

One area of future work is to develop split-merge proposals to further improve mixing rates for high-dimensional data. Although the block initialization of the time series helps with the issue of splitting merged behaviors, it does not fully solve the problem and cannot be relied upon in datasets with more irregular switching patterns than the MoCap data we considered. Additionally, splitting a single true behavior into multiple estimated behaviors often occurred. The root of the splitting issue is two-fold. One is due to the mixing rate of the sampler. The second, unlike in the case of merging behaviors, is due to modeling issues. Our model assumes that the dynamic behavior parameters (i.e., the VAR process parameters) are identical between time series and do not change over time. This assumption can be problematic in grouping related dynamic behaviors, and might be addressed via hierarchical models of behaviors or by ideas similar to those of the \emph{dependent Dirchlet process}~\citep{MacEachern:99,Griffin:06} that allows for time-varying parameters.

Overall, the MoCap results appeared to be fairly robust to examples of only slightly dissimilar behaviors (e.g., squatting to different levels, twisting at different rates, etc.) However, in cases such as the running motion where only portions of the body moved in the same way while others did not, we tended to split the behavior group. This observation motivates examination of \emph{local partition processes}~\citep{Dunson:09,Dunson:09b} rather than \emph{global partition processes}. That is, our current model assumes that the grouping of observations into behavior categories occurs along all components of the observation vector rather than just a portion (e.g., lower body measurements.) Allowing for greater flexibility in the grouping of observation vectors becomes increasingly important in high dimensions.

\section*{Acknowledgments}
This work was supported in part by MURIs funded through AFOSR Grant FA9550-06-1-0324, ARO Grant W911NF-06-1-0076, and ONR Grant N00014-11-1-0688, and by AFOSR under Grant FA9559-08-1-0180 and Grant FA9550-10-1-0501.  A preliminary version of this work (without detailed development or analysis) was first presented at a conference~\citep{Fox:NIPS09}.

\appendix

\section{Appendix A: Sum-Product Algorithm for the AR-HMM}
\label{app:sumprod}

A variant of the sum-product algorithm applied specifically to the chain graph of the AR-HMM provides an efficient method for computing the likelihood of the data marginalizing the latent HMM mode sequence.  For the BP-AR-HMM of this paper, we compute the likelihood of each time series based on a fixed feature matrix assignment, which reduces the joint model to a finite collection of finite-dimensional AR-HMMs, each of which is described by its set of feature-constrained transition distributions along with the shared library of VAR parameters $\theta_k = \{\BF{A}_k,\Sigma_k\}$. The derivations provided in this appendix directly follow those for the standard HMM~\citep{Rabiner:89}.  First, we define a set of \emph{forward messages}
\begin{align}
\alpha_t(z_t) \triangleq p(\BF{y}_1,\dots,\BF{y}_t,z_t),
\end{align}
which satisfy the recursion
\begin{align}
\alpha_{t+1}(z_{t+1}) 
&= p(\BF{y}_{t+1}\mid z_{t+1},\tilde{\BF{y}}_{t+1})\sum_{z_t}p(\BF{y}_1,\dots,\BF{y}_t \mid z_t)p(z_{t+1}\mid z_t)p(z_t)\\
&= p(\BF{y}_{t+1}\mid z_{t+1},\tilde{\BF{y}}_{t+1})\sum_{z_t}\alpha_t(z_t)p(z_{t+1}\mid z_t)\\
&= \mathcal{N}(\BF{y}_{t+1};\BF{A}_{z_{t+1}}\tilde{\BF{y}}_{t+1},\Sigma_{z_{t+1}})\sum_{z_t}\alpha_t(z_t)\pi_{z_t}(z_{t+1}).
\end{align}
The messages are initialized as
\begin{align}
\alpha_{1}(z_{1}) &= p(\BF{y}_1,\tilde{\BF{y}}_1,z_1) = \mathcal{N}(\BF{y}_{1};\BF{A}_{z_{1}}\tilde{\BF{y}}_{1},\Sigma_{z_{1}})\pi_0(z_1).
\end{align}

After running the recursion from $t=1,\dots,T$, the desired likelihood is simply computed by summing over the components of the forward message at time $T$:
\begin{align}
	p(\BF{y}_1,\dots,\BF{y}_T) = \sum_{z_T}\alpha_T(z_T).
\end{align}

Note that for the BP-AR-HMM, at each step the forward message for time series $i$ is computed by summing $z^{(i)}_t$ over the finite collection of possible HMM mode indices specified by that time series's feature vector $\fset{i}$.

\section{Appendix B: Acceptance Ratio for Birth-Death Proposal}\label{app:birthdeath}
Let us first consider a birth move in which we propose a transition from $n_i$ to $n_i+1$ unique features for time series $i$. As dictated by Eq.~\eqref{eqn:parameter_proposals}, the first $n_i$ proposed components of $\BF{\theta}_{+}'$ and $\BF{\eta}_{+}'$ are equal to the previous parameters associated with those $n_i$ features. Namely, $\theta_{+,k}'=\theta_{+,k}$ and $\eta_{+,k}'=\eta_{+,k}$ for all $k \in \{1,\dots,n_i\}$. The difference between the proposed and previous parameters arises from the fact that $\BF{\theta}_{+}'$ and $\BF{\eta}_{+}'$ contain an additional component $\theta_{+,n_i+1}'$ and $\eta_{+,n_i+1}'$, respectively, drawn from the prior distributions on these parameter spaces. Then, the acceptance ratio is given by
\begin{align}
	&r(\fset{+i}',\BF{\theta}_{+}',\BF{\eta}_{+}' \mid \fset{+i},\BF{\theta}_{+},\BF{\eta}_{+})\nonumber\\
	&\hspace{0.1in}=\frac{p(\symsubsupB{y}{1:T_i}{i}\mid [\fset{-i} \, \fset{+i}'],\theta_{1:K_+^{-i}},\BF{\theta}_{+}',\BF{\eta}^{(i)}, \BF{\eta}_{+}')p(\fset{+i}')p(\BF{\theta}_{+}')p(\BF{\eta}_{+}')} {p(\symsubsupB{y}{1:T_i}{i}\mid [\fset{-i} \, \fset{+i}],\theta_{1:K_+^{-i}},\BF{\theta}_{+},\BF{\eta}^{(i)},\BF{\eta}_{+} )p(\fset{+i})p(\BF{\theta}_{+})p(\BF{\eta}_{+})}\nonumber\\
	&\hspace{1in}\cdot \frac{q_f(\fset{+i} \mid \fset{+i}') q_{\theta}(\BF{\theta}_{+} \mid \fset{+i},\fset{+i}',\BF{\theta}_{+}') q_{\eta}(\BF{\eta}_{+} \mid \fset{+i}, \fset{+i}', \BF{\eta}_{+}')}{q_f(\fset{+i}' \mid \fset{+i}) q_{\theta}(\BF{\theta}_{+}' \mid \fset{+i}',\fset{+i},\BF{\theta}_{+}) q_{\eta}(\BF{\eta}_{+}' \mid \fset{+i}', \fset{+i}, \BF{\eta}_{+})} 
\end{align}
Noting that each component of the parameter vector $\BF{\theta}_+$ and $\BF{\eta}_+$ is drawn i.i.d., and plugging in the appropriate definitions for the proposal distributions, we have
\begin{align}
	&r(\fset{+i}',\BF{\theta}_{+}',\BF{\eta}_{+}' \mid \fset{+i},\BF{\theta}_{+},\BF{\eta}_{+})\nonumber\\
	&\hspace{0.1in}= \frac{p(\symsubsupB{y}{1:T_i}{i}\mid [\fset{-i} \, \fset{+i}'],\theta_{1:K_+^{-i}},\BF{\theta}_{+}',\BF{\eta}^{(i)}, \BF{\eta}_{+}')\mbox{Poisson}(n_i+1;\alpha/N)\prod_{k=1}^{n_i+1}p(\theta_{+,k}')p(\eta_{+,k}')} {p(\symsubsupB{y}{1:T_i}{i}\mid [\fset{-i} \, \fset{+i}],\theta_{1:K_+^{-i}},\BF{\theta}_{+},\BF{\eta}^{(i)},\BF{\eta}_{+} )\mbox{Poisson}(n_i;\alpha/N)\prod_{k=1}^{n_i}p(\theta_{+,k})p(\eta_{+,k})}\nonumber\\
	&\hspace{1in}\cdot \frac{q_f(n_i\leftarrow n_i+1) \prod_{k=1}^{n_i}\delta_{\theta_{+,k}'}(\theta_{+,k})\delta_{\eta_{+,k}'}(\eta_{+,k})} {q_f(n_i+1\leftarrow n_i) p(\theta_{+,n_i+1}')p(\eta_{+,n_i+1}') \prod_{k=1}^{n_i}\delta_{\theta_{+,k}}(\theta_{+,k}')\delta_{\eta_{+,k}}(\eta_{+,k}')}. 
\end{align}
We use the notation $q_f(k\leftarrow j)$ to denote the proposal probability of transitioning from $j$ to $k$ unique features. Using the fact that $\theta_{+,k}'=\theta_{+,k}\in \theta_{1:K_+}$ and $\eta_{+,k}'=\eta_{+,k}\in \BF{\eta}^{(i)}$ for all $k \in \{1,\dots,n_i\}$, we can simplify the acceptance ratio to:
\begin{align}
	\frac{p(\symsubsupB{y}{1:T_i}{i}\mid [\fset{-i} \, \fset{+i}'],\theta_{1:K_+},\theta_{+,n_i+1}',\BF{\eta}^{(i)}, \eta_{+,n_i+1}')\mbox{Poisson}(n_i+1;\alpha/N)q_f(n_i\leftarrow n_i+1)} {p(\symsubsupB{y}{1:T_i}{i}\mid [\fset{-i} \, \fset{+i}],\theta_{1:K_+},\BF{\eta}^{(i)}) \mbox{Poisson}(n_i;\alpha/N)q_f(n_i+1\leftarrow n_i)}. 
\end{align}
The derivation of the acceptance ratio for a death move follows similarly.

\section{Appendix C: Acceptance Ratio for Transition Parameters}
\label{app:transparams}
Since the proposal distributions for $\gamma$ and $\kappa$ use fixed variance $\sigma_\gamma^2$ or $\sigma_\kappa^2$, and mean equal to the current hyperparameter value, we have
\begin{align}
	q_\gamma(\cdot \mid \gamma) = \mbox{Gamma}\left(\frac{\gamma^2}{\sigma_\gamma^2},\frac{\gamma}{\sigma_\gamma^2}\right) \hspace{0.25in} q_\kappa(\cdot \mid \kappa) = \mbox{Gamma}\left(\frac{\kappa^2}{\sigma_\kappa^2},\frac{\kappa}{\sigma_\kappa^2}\right). \label{eqn:GammaProposals} 
\end{align}
Let $\BF{\pi} = \{\pi_j^{(i)}\}$. To update $\gamma$ given $\kappa$, the acceptance probability is $\min\{r(\gamma' \mid \gamma),1\}$ with acceptance ratio
\begin{align}
	r(\gamma' \mid \gamma) = \frac{p(\BF{\pi}\mid \gamma',\kappa,\BF{F})p(\gamma'\mid a_\gamma,b_\gamma)q(\gamma \mid \gamma',\sigma^2_{\gamma})}{p(\BF{\pi}\mid \gamma,\kappa,\BF{F})p(\gamma\mid a_\gamma,b_\gamma)q(\gamma' \mid \gamma,\sigma^2_{\gamma})}, 
\end{align}
Recalling the definition of $\symsubsup{\tilde{\pi}}{j}{i}$ from Eq.~\eqref{eqn:DirPrior} and that $K_i = \sum_k f_{ik}$, the likelihood term may be written as
\begin{align}
	f(\gamma) \triangleq p(\BF{\pi}\mid \gamma,\kappa,\BF{F}) = \prod_i \prod_{k=1}^{K_i} \left\{\frac{\Gamma(\gamma K_i + \kappa)}{\left(\prod_{j=1}^{K_i-1} \Gamma(\gamma)\right)\Gamma(\gamma+\kappa)} \prod_{j=1}^{K_i} \tilde{\pi}_{kj}^{(i)^{\gamma+\kappa\delta(k,j)-1}}\right\}. 
\end{align}
The ratio of the prior distributions reduces to
\begin{align}
	\frac{p(\gamma' \mid a_\alpha,b_\alpha)}{p(\gamma \mid a_\alpha,b_\alpha)} = \left(\frac{\gamma'}{\gamma}\right)^{a_\gamma-1} \exp\{-(\gamma' - \gamma)b_\gamma\}. 
\end{align}
Letting $\vartheta=\gamma^2/\sigma_{\gamma}^2$ and $\vartheta'=\gamma'^2/\sigma_{\gamma}^2$, the ratio of the proposal distributions reduces to
\begin{align}
	\frac{q(\gamma \mid \gamma',\sigma_\gamma^2)}{q(\gamma' \mid \gamma,\sigma_\gamma^2)} = \frac{\frac{(\gamma'/\sigma_\gamma^2)^{\vartheta'}}{\Gamma(\vartheta')}\gamma^{\vartheta'-1}\exp\{-\gamma \frac{\gamma'}{\sigma_\gamma^2}\}}{\frac{(\gamma/\sigma_\gamma^2)^\vartheta}{\Gamma(\vartheta)}\gamma'^{\vartheta-1}\exp\{-\gamma' \frac{\gamma}{\sigma_\gamma^2}\}} = \frac{\Gamma(\vartheta)\gamma^{\vartheta'-\vartheta-1}}{\Gamma(\vartheta')\gamma'^{\vartheta-\vartheta'-1}}\sigma_\gamma^{2(\vartheta-\vartheta')}. 
\end{align}
Our acceptance ratio can then be compactly written as
\begin{align}
	r(\gamma' \mid \gamma) = \frac{f(\gamma')\Gamma(\vartheta)\gamma^{\vartheta'-\vartheta-a_\gamma}}{f(\gamma)\Gamma(\vartheta')\gamma'^{\vartheta-\vartheta'-a_\gamma}} \exp\{-(\gamma' - \gamma)b_\gamma\} \sigma_\gamma^{2(\vartheta-\vartheta')}. 
\end{align}

The Metropolis-Hastings sub-step for sampling $\kappa$ given $\gamma$ follows similarly. In this case, however, the likelihood terms simplifies to
\begin{align}
	f(\kappa) \triangleq \prod_i \frac{\Gamma(\gamma K_i + \kappa)^{K_i}}{\Gamma(\gamma+\kappa)^{K_i}} \prod_{j=1}^{K_i} \tilde{\pi}_{jj}^{(i)^{\gamma+\kappa-1}} \propto p(\BF{\pi}\mid \gamma,\kappa,\BF{F}). 
\end{align}

\section{Appendix D: BP-AR-HMM MCMC Algorithm}
\label{app:alg}
The overall MCMC sampler for the BP-AR-HMM is outlined in Algorithm~\ref{alg:IBPARHMMsampler}.  Note that Algorithm~\ref{alg:IBPARHMMzsampler} is embedded within Algorithm~\ref{alg:IBPARHMMsampler}.
\begin{algorithm}[p!] \vspace*{-1pt} \hspace*{-6pt} 
	\begin{flushleft}
		\vspace{0pt} Given a previous set of time-series-specific transition variables $\{\BF{\eta}^{(i)}\}^{(n-1)}$, the dynamic parameters $\{\BF{A}_k,\Sigma_k\}^{(n-1)}$, and features $\BF{F}^{(n-1)}$: 
		\begin{enumerate}
			\algtop 
			\item Set $\{\BF{\eta}^{(i)}\}=\{\BF{\eta}^{(i)}\}^{(n-1)}$, $\{\BF{A}_k,\Sigma_k\} = \{\BF{A}_k,\Sigma_k\}^{(n-1)}$, and $\BF{F}=\BF{F}^{(n-1)}$. \vspace{-0.05in} 
			\item From the feature matrix $\BF{F}$, create count vector $\BF{m} = [
			\begin{array}{cccc}
				m_1 & m_2 & \dots & m_{K_+}
			\end{array}
			]$, with $m_k$ representing the number of time series possessing feature $k$. \vspace{-0.05in} 
			\item For each $i \in \{1,\dots,N\}$, sample features as follows: 
			\begin{enumerate}
				\algtop 
				\item Set $\BF{m}^{-i} = \BF{m} - \fset{i}$, and reorder columns of $\BF{F}$ so that the $K_+^{-i}$ columns with $m_k^{-i}>0$ appear first. Appropriately relabel indices of $\{\BF{A}_k,\Sigma_k\}$ and $\{\BF{\eta}^{(i)}\}$. 
				\item For each shared feature $k \in \{1,\dots,K_+^{-i}\}$, set $f=f_{ik}$ and: 
				\begin{enumerate}
					\algtop \vspace{0.05in} 
					\item Consider $f_{ik} \in \{0,1\}$ and: 
					\begin{enumerate}
						\algtop \vspace{0.05in} 
						\item Create feature-constrained transition distributions: \vspace{0.05in} 
						\begin{align*}
							\symsubsup{\pi}{j}{i} \propto [
							\begin{array}{cccc}
								\symsubsup{\eta}{j1}{i} & \symsubsup{\eta}{j2}{i} & \dots & \symsubsup{\eta}{j{K_+}}{i}\; 
							\end{array}
							] \otimes \fset{i} 
						\end{align*}
						\item Compute likelihood $\ell_{f_{ik}}\left(\symsubsupB{y}{1:T_i}{i}\right) \triangleq p\left(\symsubsupB{y}{1:T_i}{i}\mid \BF{\pi}^{(i)},\{\BF{A}_k,\Sigma_k\}\right)$ using a variant of the sum-product algorithm described in Appendix~\ref{app:sumprod}. 
					\end{enumerate}
					\item Compute 
					\begin{align*}
						\rho^* = \frac{m_k^{-i}}{N-m_k^{-i}}\cdot\frac{\ell_1\left(\symsubsupB{y}{1:T_i}{i}\right)}{\ell_0\left(\symsubsupB{y}{1:T_i}{i}\right)} \hspace{0.1in} \mbox{and set} \hspace{0.1in} \rho(\bar{f}\mid f) = \left\{ 
						\begin{array}{ll}
							\min\{\rho^*,1\}, & f=0; \\
							\min\{1/\rho^*,1\}, & f=1. 
						\end{array}
						\right. 
					\end{align*}
					\item Sample $f_{ik} \sim \rho(\bar{f}\mid f)\delta(f_{ik},\bar{f}) + (1-\rho(\bar{f}\mid f))\delta(f_{ik},f)$. 
				\end{enumerate}
				\item Let \fset{i}' = \fset{i} and calculate the number of unique features $n_i = K_+ - K_+^{-i}$. 
				\begin{enumerate}
					\algtop \vspace{0.05in} 
					\item Propose a birth or death move, each with probability 0.5. 
					\begin{itemize}
						\item Birth: sample $\{\theta_{+,n_i+1}',\eta_{+,n_i+1}\}$ from their priors and set $f_{i,n_i+1}' = 1$, $n_i' = n_i+1$. 
						\item Death: sample $\ell \sim \mbox{uniform}[K_+^{-i}+1:K_+]$ and set $f_{i\ell}'=0$, $n_i' = n_i - 1$. 
					\end{itemize}
					\item Compute likelihoods $\ell_{\fset{i}}\left(\symsubsupB{y}{1:T_i}{i}\right)$ and $\ell_{\fset{i}'}\left(\symsubsupB{y}{1:T_i}{i}\right)$ of data under the previous and proposed models, respectively. 
					\item Keep ($\zeta=1$) or discard ($\zeta=0$) proposed model by sampling:					
					\begin{align*}
						\hspace{-0.1in}\zeta \sim \mbox{Ber}(\rho) \hspace{0.15in} \rho = \min \left\{\frac{\ell_{\fset{i}}\left(\symsubsupB{y}{1:T_i}{i}\right)\mbox{Poisson}(n_i'\mid \frac{\alpha}{N})q_f(n_i \leftarrow n_i')}{\ell_{\fset{i}'}\left(\symsubsupB{y}{1:T_i}{i}\right)\mbox{Poisson}(n_i\mid \frac{\alpha}{N})q_f(n_i' \leftarrow n_i)},1\right\}. 
					\end{align*}
				\end{enumerate}
				\item Set $\BF{m} = \BF{m}^{-i}+\fset{i}$. Remove columns for which $m_k=0$, and appropriately redefine the dynamic parameters $\{\BF{A}_k,\Sigma_k\}$ and transition variables $\{\BF{\eta}^{(i)}\}$. 
			\end{enumerate}
			\item Resample dynamic parameters $\{\BF{A}_k,\Sigma_k\}$ and transition variables $\{\BF{\eta}^{(i)}\}$ using the auxiliary variable sampler of Algorithm~\ref{alg:IBPARHMMzsampler}. 
			\item Fix $\{\BF{\eta}^{(i)}\}^{(n)} = \{\BF{\eta}^{(i)}\}$, $\{\BF{A}_k,\Sigma_k\}^{(n)}=\{\BF{A}_k,\Sigma_k\}$, and $\BF{F}^{(n)}=\BF{F}$. \algend 
		\end{enumerate}
	\end{flushleft}
	\caption{BP-AR-HMM MCMC sampler.} \label{alg:IBPARHMMsampler} 
\end{algorithm}
\begin{algorithm}[htbp] \vspace*{-1pt} \hspace*{-6pt} 
	\begin{flushleft}
		\vspace{0pt} Given the feature-restricted transition distributions $\BF{\pi}^{(i)}$ and dynamic parameters $\{\BF{A}_k,\Sigma_k\}$, update the parameters as follows: 
		\begin{enumerate}
			\item For each $i \in \{1,\dots,N\}$: 
			\begin{enumerate}
				\item Block sample $\symsubsup{z}{1:T_i}{i}$ as follows: 
				\begin{enumerate}
					\algtop 
					\item For each $k\in \{1,\dots,K_+\}$, initialize messages to $\symsubsup{m}{T+1,T}{i}(k)=1$.
					
					\item For each $t \in \{T_i,\ldots,1\}$ and $k\in \{1,\dots,K_+\}$, compute 
					\begin{equation*}
						\symsubsup{m}{t,t-1}{i}(k) = \sum_{j=1}^K \symsubsup{\pi}{k}{i}(j)\mathcal{N}\left(\symsubsupB{y}{t}{i}; \BF{A}_{j}\symsubsupB{\tilde{y}}{t}{i},\Sigma_{j}\right)\symsubsup{m}{t+1,t}{i}(j). 
					\end{equation*}
					\vspace{-4pt} \algend 
					\item Working sequentially forward in time, and starting with transitions counts $\symsubsup{n}{j k}{i}=0$: 
					\begin{enumerate}
						\algtop 
						\item Sample a mode assignment $\symsubsup{z}{t}{i}$ as: 
						\begin{align*}
							\symsubsup{z}{t}{i} &\sim \sum_{k=1}^{K_+} \symsubsup{\pi}{\symsubsup{z}{t-1}{i}}{i}(k)\mathcal{N}\left(\symsubsupB{y}{t}{i}; \BF{A}_{k}\symsubsupB{\tilde{y}}{t}{i},\Sigma_{k}\right)\symsubsup{m}{t+1,t}{i}(k)\delta\left(\symsubsup{z}{t}{i},k\right). 
						\end{align*}
						\item Increment $\symsubsup{n}{\symsubsup{z}{t-1}{i}\symsubsup{z}{t}{i}}{i}$. \algend 
					\end{enumerate}
				\end{enumerate}
				Note that $\symsubsup{\pi}{j}{i}(k)$ is zero for any $k$ such that $f_{ik} = 0$, implying that $\symsubsup{z}{t}{i}=k$ will never be sampled (as desired). Considering all $K_+$ indices simply allows for efficient matrix implementation. 
				\item For each $(j,k) \in \{1,\dots,K_+\} \times \{1,\dots,K_+\}$, sample 
				\begin{align*}
					\symsubsup{\eta}{jk}{i}\mid \gamma \sim \mbox{Gamma}(1,\gamma+\kappa\delta(j,k)+\symsubsup{n}{jk}{i}). 
				\end{align*}
			\end{enumerate}
			\item For each $k \in \{1,\dots,K_+\}$: 
			\begin{enumerate}
				\item Form $\BF{Y}_k = \{\symsubsupB{y}{t}{i} | \symsubsup{z}{t}{i} = k\}$ and $\BF{\tilde{Y}}_k = \{\symsubsupB{\tilde{y}}{t}{i} | \symsubsup{z}{t}{i} = k\}$ and compute $\symsubsup{S}{\tilde{y}\tilde{y}}{k}$, $\symsubsup{S}{y\tilde{y}}{k}$, $\symsubsup{S}{yy}{k}$, and $\symsubsup{S}{y\mid \tilde{y}}{k}$ as in Eq.~\eqref{eqn:Sk}. 
				\item Sample dynamic parameters: 
				\begin{align*}
					\Sigma_k &\sim \mbox{IW}\left(\sum_{i=1}^N \symsubsup{n}{k\cdot}{i} + n_0, \symsubsup{S}{y\mid \tilde{y}}{k} + S_0\right)\\
					\BF{A}_k \mid \Sigma_k &\sim \MN{\BF{A}_k}{\symsubsup{S}{y\tilde{y}}{k}S_{\tilde{y}\tilde{y}}^{-(k)}}{\Sigma_k}{\symsubsup{S}{\tilde{y}\tilde{y}}{k}}. 
				\end{align*}
			\end{enumerate}
		\end{enumerate}
	\end{flushleft}
	\caption{BP-AR-HMM auxiliary variable sampler for updating transition and dynamic parameters.} \label{alg:IBPARHMMzsampler} 
\end{algorithm}

\bibliographystyle{chicago}
\bibliography{../../Bibliography/Bibliography_BPARHMM}

\end{document}